\documentclass[aps,prb,reprint,twocolumn]{revtex4}

\usepackage{commath}
\usepackage{amsmath}
\usepackage{graphicx}
\usepackage{hyperref}
\usepackage{txfonts}

\draft 

\begin{document}


\title{Decomposition into Propagating and Evanescent Modes of Graphene Ribbons}


\author{Hai-Yao Deng}
\author{Katsunori Wakabayashi}
\email[Corresponding author:~]{WAKABAYASHI.Katsunori@nims.go.jp}
\affiliation{International Center for Materials Nanoarchitechtonics (WPI-MANA),
National Institute for Materials Science (NIMS), Namiki 1-1, Tsukuba
305-0044, Japan}  

\date{\today}

\begin{abstract}
Bulk modes (BM) are basis solutions to the
 Schr\"{o}dinger equation and they are useful in a number of physical
 problems. In the present work, we establish a complete set of BMs for
 graphene ribbons at arbitrary energy. We derive analytical expressions for
 these modes and systematically classify them into propagating or
 evanescent mode. We also demonstrate their uses in efficient electronic transport
 simulations of graphene-based electronic devices within both the
 mode-matching method and the Green's function framework. Explicit
 constructions of Green's functions for infinite and semi-infinite
 graphene ribbons are presented.  
\end{abstract}


\maketitle 

\section{Introduction}
\label{section:1}
Graphene, which is an atomically thick carbon sheet, and its nanostructures such
as graphene nanoribbons continue to attract immense interest in the past
decade due to their peculiar
properties.~\cite{novoselov2004,geim2005,novoselov2005b,fujita1996,nakada1996,wakabayashi1999,wakabayashi1998jpsj,kane2005,wakabayashi2007,white2007,wakabayashi2009a,wakabayashi2009b,geim2007rise,neto2009electronic,wakabayashi2012nanoscale}
Numerous electronic devices~\cite{raza2012graphene} based on them, such as p-n junctions,
field effect transistors and memory devices as well as electro-optical wave guides, have been studied
extensively from both the
theoretical~\cite{cheianov2006,abanin2007,deng2012,gunlycke2007,areshkin2007}
and experimental point of
view.~\cite{ozyilmaz2007,huard2007,williams2007}  

Theoretically, quantum transport simulations play an important part in understanding the
behaviors of nano-devices.~\cite{londergan1999binding} Such simulations
are generally based on Landauer-B\"uttiker
picture~\cite{landauer1957spatial,landauer1970electrical,buttiker1986}
and can be implemented using either mode-matching~\cite{ando1991quantum,khomyakov2008,sorensen2009efficient}
or Green's function approaches.~\cite{economou1981,fisher1981relation,datta1997electronic} As shown by
Khomyakov \textit{et al.},~\cite{khomyakov2005conductance} these approaches are
equivalent and their key quantities can be expressed in terms of the
bulk modes (BMs), which are basis solutions (without satisfying all
boundary conditions) to the Schr\"{o}dinger equation and hence
characteristic of the underlying Hamiltonian. However, a systematic
exposure of such BMs for graphene structures has so far been wanting in
the literature.    

In the present paper, we systematically derive and classify the BMs (as
either propagating or evanescent depending on their far-field behaviors) for
graphene and graphene ribbons at arbitrary energy. Evanescent modes are known
as the exponentially decaying or growing modes, which are induced at the edge surface or scattering center. 
We present those results for both armchair and
zigzag ribbons. For graphene and armchair graphene ribbons (AGRs),
the transverse (perpendicular to ribbon direction) and longitudinal
(along ribbon direction) electronic motions are decoupled and
hence analytical expressions for BMs can be obtained at arbitrary energy.~\cite{wurm2009interfaces,iyengar2008} For
zigzag graphene ribbons (ZGRs), these motions are not decoupled and
analytical expressions are generally not available.~\cite{brey2006,wakabayashi1999} However, we derive a
simple polynomial equation, which can be easily solved numerically, for
locating the BMs of ZGRs. At low energies, simple analytical expressions
are found to this equation.  

A complete set of BMs is suitable for solving electron
scattering problem of graphene
nanostructures.~\cite{ando1991quantum,sorensen2009efficient,khomyakov2005conductance,beenakker1997}
We demonstrate this by studying a graphene point contact in the
mode-matching approach and by explicit construction of lattice Green's
functions, which are essential in transport simulations of nano-devices,~\cite{datta1997electronic,yamamoto2009,rycerz2007,wakabayashi2000} for infinite and semi-infinite 
ribbons.

This paper is organized as follows. In the next section, we introduce
the representation and derive the BMs for graphene. Then, in
section~\ref{section:3}, we construct the BMs for ZGRs and thoroughly
analyze their properties. Section~\ref{section:4} is devoted to various
applications as mentioned above. In Appendix A, we present results for
AGRs. In Appendix B, we derive a useful formula for calculating group
velocity of any mode.    

\section{Bulk Modes of Graphene}
\label{section:2}
\begin{figure}
\includegraphics[width=0.45\textwidth]{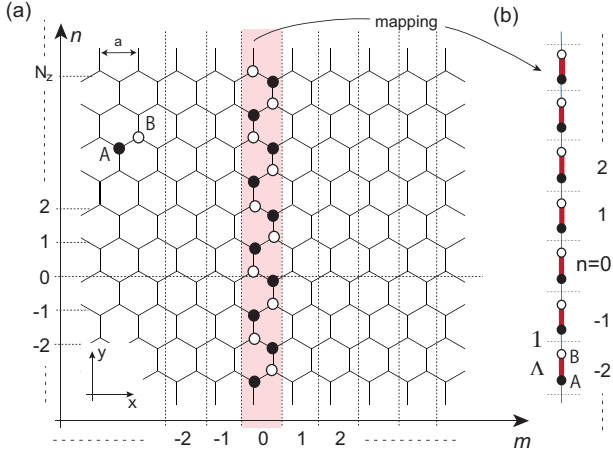}
\caption{(a) Schematic of graphene lattice. In the $x$-direction, the lattice is viewed as
 repetition of the supercell (numbered by $m$) indicated by the shaded rectangle. 
In the $y$-direction, the lattice is a collection of zigzag
 chains, which are labeled by $n$. A supercell extends from $n=-\infty$ to $n=\infty$. 
Each lattice site is specified as $(m,n,\nu)$, where $\nu=A,B$
 refers to the sublattices. 
(b) By a transformation expressed in Eq.(\ref{M}), the supercell is
 transformed into a diatomic chain. 
The effective hopping energies between adjacent sites are either
 $\Lambda$ or $1$, depending on whether the sites sit in the same zigzag
 chain or not.} 
\label{figure:F1}
\end{figure}

Graphene has a honeycomb lattice structure of carbon atoms as shown in Fig.~\ref{figure:F1} (a). 
The $x$ ($y$)-axis is taken along (perpendicular to) zigzag
chains. For the purpose of this paper, we construct a supercell for 
the translational operation along $x$-axis, which is indicated as
the rectangle region containing a single armchair chain in the figure.
Since the honeycomb lattice is AB-bipartite, the atomic site 
on the $n$-th zigzag chain in the $m$-th supercell can be
specified by three indices, $(m,n,\nu)$, where $\nu=$A, B. For
convenience, we define the sublattice index $\bar{\nu}$ to have  
the relations $\bar{\rm A}=$B and $\bar{\rm B}=$A. Throughout this
paper, we choose the unit of length to be the lattice constant $a$,
which is 0.142 nm.

We employ the nearest-neighbor tight-binding model to describe the
electronic states of graphene. The Hamiltonian is written as
\begin{equation}
H = -\gamma_0\sum_{\langle m,n,\nu, m^\prime,n^\prime \rangle}  |m,n,\nu\rangle\langle m^\prime,n^\prime,\bar{\nu}|,
\label{H}
\end{equation}
where the summation is taken only for the pairs of nearest-neighbor
carbon sites.
$\gamma_0$ is the nearest-neighbor hopping energy which is approximately
$2.7$ eV. 
The Schr\"odinger equation for a given energy $E$ is written as 
\begin{equation}
\quad H |F\rangle = E |F\rangle
\label{F_sch}
\end{equation}  
with a generic solution $|F\rangle$, which can be decomposed as
\begin{equation}
|F\rangle = \sum_{m,n,\nu} F_\nu(m,n)|m,n,\nu\rangle.
\label{F}
\end{equation}  
The equation of motion for $F_\nu(m,n)$ is given by
\begin{eqnarray}
\varepsilon F_\nu(m,n) = \sum_{m', n'} F_{\bar{\nu}}(m',n'),
\label{em}
\end{eqnarray}
where $\varepsilon = -E/\gamma_0$ is the dimensionless energy. 
Thus, $F_\nu(m,n)$ is given by the summation of its 
nearest-neighbors' $F_{\bar{\nu}}(m',n')$. 

Since Eq.~(\ref{em}) is linear, any $F_{\nu}(m,n)$ can be resolved 
by a complete set of BMs. To obtain the BMs at arbitrary energy, we make use of the fact that
the coefficients in Eq.~(\ref{em}) are independent of the supercell index $m$. Therefore, 
Bloch-esque ansatz~\cite{ando1991quantum,khomyakov2005conductance} can be applied to Eq.~(\ref{em}), 
according to which $F_\nu(m,n)$ is related to $F_\nu(m-1,n)$ by the Bloch factor (a complex parameter) $\lambda$,
i.e., 
\begin{equation}
F_\nu(m,n) = \lambda F_\nu(m-1,n) = \lambda^m F_\nu(0,n).
\label{blochn}
\end{equation} 
Note that within a supercell the relative position between the $n$-th
$A$ site and the $n$-th $B$ site depends on the parity of $n$. To remove
this dependence, we adopt the following transformation (with coordinates
chosen in such a way that in the zigzag chain $n=1$ the $A$ site is to
the left of the  $B$ site, see Fig.~\ref{figure:F1}; otherwise, the
following $\hat{M}$ matrix has be to replaced by its inverse), 
\begin{equation}
\hat{f}(n)=\hat{M}^{-1} \hat{F}(0,n), \quad \hat{M} = 
\begin{pmatrix}
\lambda^{\frac{(-1)^{n}}{4}} & 0 \\
0 & \lambda^{-\frac{(-1)^{n}}{4}}
\end{pmatrix}
\label{M}
\end{equation}
where we have defined the spinors 
\begin{equation}
\hat{F}(m,n) = 
\begin{pmatrix}
F_A(m,n) \\
F_B(m,n)
\end{pmatrix},\quad
\hat{f}(n) = 
\begin{pmatrix}
f_A(n)\\
f_B(n)
\end{pmatrix}.
\label{f}
\end{equation}
After this transformation, 
the equation of motion for graphene is reduced to 
that for a diatomic linear chain as shown in Fig.~\ref{figure:F1}(b). 
Substituting Eqs.~(\ref{blochn}) and (\ref{M}) in Eq.~(\ref{em}), we
arrive at 
\begin{equation}
\begin{array}{c}
\varepsilon f_{A}(n) = f_{B}(n-1) + \Lambda f_{B}(n), \\
\varepsilon f_{B}(n) = f_{A}(n+1) + \Lambda f_{A}(n),
\end{array}
\label{eom}
\end{equation}
where 
\begin{equation}
\Lambda = \sqrt{\lambda} + \sqrt{1/\lambda}.  
\label{lamdatoLambda}
\end{equation}
Now we can apply a similar Bloch ansatz regarding $n$, namely,
\begin{equation}
f_{\nu}(n) = \sigma^{n} f_{\nu},
\label{blochl}
\end{equation}
where $f_{\nu}\coloneqq f_{\nu}(n=0)$ and $\sigma$ is another Bloch factor. Inserting Eq.~(\ref{blochl}) in (\ref{eom}), we get
\begin{equation}
\varepsilon 
\begin{pmatrix}
f_A \\
f_B
\end{pmatrix} = 
\begin{pmatrix}
0 &  \Lambda + \frac{1}{\sigma}\\
\Lambda + \sigma  & 0
\end{pmatrix}
\begin{pmatrix}
f_A \\
f_B
\end{pmatrix},
\label{se}
\end{equation}
which is solved to yield
\begin{equation}
\begin{pmatrix}
f_A \\
f_B
\end{pmatrix} = \mathcal{N}
\begin{pmatrix}
s\sqrt{\Lambda+\sigma^{-1}} \\
\sqrt{\Lambda + \sigma} 
\end{pmatrix}
\eqqcolon
\hat{f}_{\lambda\sigma s},
\label{ff} 
\end{equation}
and 
\begin{equation}
\varepsilon^2 = (\Lambda + \sigma) (\Lambda + \frac{1}{\sigma}).
\label{E}
\end{equation}
Here $\mathcal{N}$ is a normalization factor and $s=\pm$ indicates the
sign of $\varepsilon$. Note that $\varepsilon$ is symmetric with respect
to $\lambda$ and $\lambda^{-1}$, also with respect to $\sigma$ and
$\sigma^{-1}$.
Thus, once we obtain the wavefunction for diatomic chain system for a
given energy $\varepsilon$, the BMs for graphene can be 
obtained as a function of $\lambda$, $\sigma$ and $s$ through the
following relation:
\begin{equation}
\hat{F}(m,n;\lambda,\sigma,s) = \lambda^m \sigma^n \hat{M} \hat{f}_{\lambda\sigma s}.  
\label{finalF}
\end{equation}  
Although these modes are linearly independent of each other, they are
not all orthogonal, because the square matrix in Eq.~(\ref{se}) is not
hermitian in general. 
Note that Eq.~(\ref{finalF}) is obtained independent of boundary conditions and follows
solely from the homogeneity (as embodied by the Bloch ansatz) and
linearity of Eq.~(\ref{em}). 

The boundary conditions for graphene are encoded in the translational symmetry along
$x$- and $y$-directions, which require $|\lambda|=|\sigma|=1$, i.e.,
$\lambda = e^{ik}$ and $\sigma = e^{ip}$, where $k$ and $p$ denote wave
numbers in the range $(-\pi,\pi]$. 
The energy spectrum is then obtained as 
\begin{equation}
 \varepsilon = s \sqrt{1+2 g_k \cos(p) + g_k^2},
\end{equation}
where $g_k := 2\cos(k/2)$. The celebrated Dirac points are located at
$k=\pm\frac{2\pi}{3}$ and $p=\pi$. 

\begin{figure*}
\includegraphics[width=\textwidth]{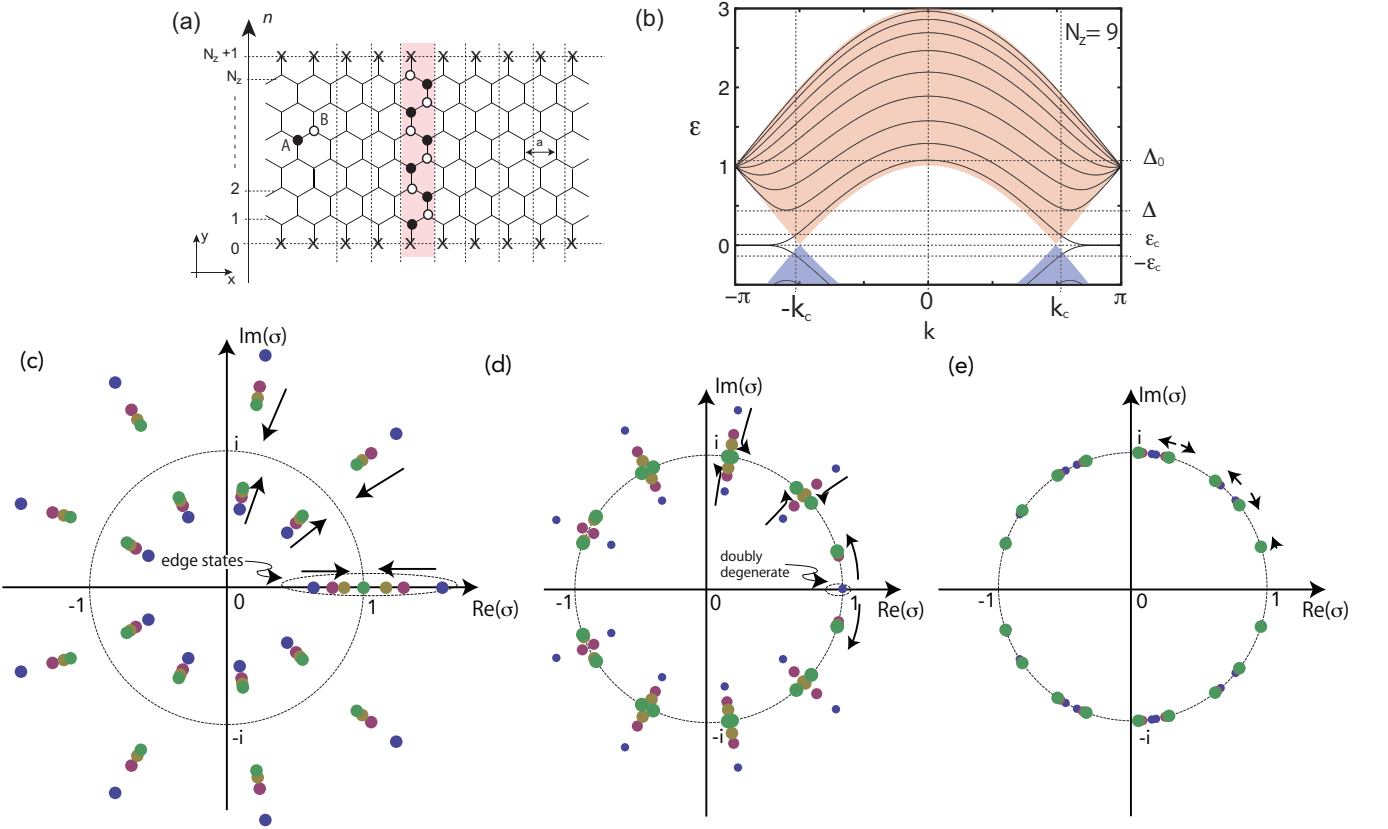}
\caption{ (a) Schematic of ZGR lattice. The symbol ``x'' means that the wave
 function should be zero at this site. 
(b) Energy band structure of ZGR for $N_z=9$. Shaded region represents
 the spectrum of graphene. (c) Distribution of roots of Eq.~(\ref{poly}) in the complex
 $\sigma$-plane for the energy region  $0<|\varepsilon|\leq\varepsilon_c$. 
In this plot, the energies are taken as $\varepsilon=0.01,0.04,0.07,0.10$.
The arrows indicate the direction of increasing energy.
None of the roots (except when $|\varepsilon|=\varepsilon_c$; note $\varepsilon_c=1/(N_z+1) = 0.1$ for $N_z=9$.) sit on the unit circle.
(d) Same as (c) for the energy region of 
 $\varepsilon_c\geq|\varepsilon|\le1$. 
In this plot, the energies are taken as $\varepsilon=0.1,0.4,0.7,1.0$.
(e) Same as (c) for the energy region of $1\le|\varepsilon|$. 
In this plot, the energies are taken as
 $\varepsilon=1.0,1.5,2.0,2.5,3.0$. Some of the circles are
overlapping because of the tiny shift.
Here all roots lie on the unit circle (dashed curve), i.e., $|\sigma|=1$ and the resulting
$p$ are real.}
\label{figure:F2}
\end{figure*}

\section{Mode Decomposition of Zigzag Graphene Ribbons}
\label{section:3}
\subsection{Derivation of basic equations}
In this section we derive a complete set of BMs for ZGRs. These BMs can be classified 
as either propagating or evanescent, as described below.
We assume that the ZGR has $N_z$ zigzag chains in the region $0<n<N_z+1$.
Let $\psi_\nu(m,n)$ denote the BMs together with its spinor
representation:
\begin{equation}
\hat{\psi}(m,n) = 
\left(
 \begin{array}{c}
\psi_{\rm A}(m,n) \\
\psi_{\rm B}(m,n)
 \end{array}
\right).
\end{equation}
The boundary conditions then require
$\psi_B(m,0)=0$ and $\psi_A(m,N_z+1)=0$. 
Since the energy $\varepsilon$ is invariant under $\sigma \rightarrow
\frac{1}{\sigma}$, $\psi_\nu(m,n)$ can be simply derived from
the linear combination of 
$\hat{F}(m,n {;\lambda,\sigma,s})$ and
$\hat{F}(m,n {;\lambda,\sigma^{-1},s})$, i.e.,
\begin{equation}
\hat{\psi}({m,n;\lambda,\sigma, s}) = \lambda^m \hat{M} (\beta \sigma^n \hat{f}_{\lambda\sigma s}-\beta'\sigma^{-n}\hat{f}_{\lambda\sigma^{-1}s}).
\label{psi}
\end{equation} 
Here $\beta$ and $\beta'$ are coefficients to be determined by the
boundary conditions. Note that the $\hat{\psi}({m,n;\lambda,\sigma, s})$ and $\hat{\psi}({m,n;\lambda,\sigma^{-1},s})$
represent the same BM. 

By imposing the boundary conditions to Eq.~(\ref{psi}), we obtain
\begin{equation}
\begin{pmatrix}
z & - \frac{1}{z} \\
\Lambda + \sigma  & -\left(\Lambda + \frac{1}{\sigma}\right)
\end{pmatrix}
\begin{pmatrix}
\beta \\
\beta'
\end{pmatrix}=0,
\label{beta}
\end{equation} 
where $z = \sigma^{N_z+1}$. Non-zero solutions exist if
\begin{equation}
z^2 = \frac{\Lambda + \sigma}{\Lambda + \frac{1}{\sigma}}.
\label{lamsig}
\end{equation} 
Together with Eq.~(\ref{E}), this equation determines the allowed values
of $(\lambda,\sigma)$ at fixed energy $\varepsilon$. 

By plugging it in Eq.~(\ref{psi}), we find
\begin{equation}
\hat{\psi}(m,n; {\lambda,\sigma, s}) = \lambda^m \hat{\Phi}(n;{\lambda,\sigma, s}), 
\label{finalpsi}
\end{equation}
where 
\begin{equation}
\hat{\Phi}(n;{\lambda,\sigma, s})\coloneqq \mathcal{N}' \hat{M}
\begin{pmatrix}
S_{N_z+1-n}(\sigma) \\
s(-1)^QS_n(\sigma)
\end{pmatrix}.
\label{Phi}
\end{equation}
Here $\mathcal{N}'$ is a normalization factor and 
\begin{equation}
S_{n}(\sigma) = \frac{1}{2 i}\left[\sigma^n -
			      \left(\frac{1}{\sigma}\right)^n\right]. 
\label{S}
\end{equation}
$Q$ is an integer dictating the parity of the mode. 
Explicitly, we have~\cite{note3} 
\begin{equation}
(-1)^Q=z^{-1}\sqrt{\frac{\Lambda+\sigma}{\Lambda+\sigma^{-1}}}.
\label{Q}
\end{equation}
See that $\sigma = \pm 1$ lead to $S_n\equiv 0$, which must be
excluded. These relations, $S_n(\sigma^{-1})=-S_n(\sigma)$,
$S_n(-\sigma)=(-)^n S_n(\sigma)$, generally hold.

\subsection{Energy spectrum and number of modes}
In this subsection we will show that the derived set of equations can
correctly reproduce the energy band structure of ZGRs.
To facilitate further analysis, we combine Eqs.~(\ref{E}) and
(\ref{lamsig}) to get a polynomial equation of degree $4N_z$ for
$\sigma$, 
\begin{equation}
\left\{\mathcal{G}_{N_z}(\sigma^2) - \frac{\sigma^{N_z}}{\varepsilon}\right\}\cdot
 \left\{\mathcal{G}_{N_z}(\sigma^2) + \frac{\sigma^{N_z}}{\varepsilon} \right\} = 0.
\label{poly}
\end{equation} 
Here $\mathcal{G}_{N_z}(x) = 1 + x + x^2 + ... + x^{N_z}$ and we have already excluded one pair of the unwanted
roots $\sigma^2=1$. 
In addition, the $\Lambda$ is derived from $\sigma$ using the following relation
\begin{equation}
\Lambda = -\frac{S_{N_z}(\sigma)}{S_{N_z+1}(\sigma)}.
\label{lam}
\end{equation}
The energy spectrum for ZGR can be obtained if we assume a conventional
Bloch phase for $\lambda=\exp(i k)$, which immediately gives
$\Lambda=2\cos(k/2)\eqqcolon g_k$. Then we arrive at
\begin{equation}
g_k = -\frac{S_{N_z}(\sigma)}{S_{N_z+1}(\sigma)}.
\label{lam2}
\end{equation}
This relation was derived by one of us and is confirmed to reproduce
the energy spectrum,~\cite{wakabayashi2012nanoscale} if we solve this equation
under the condition that $\sigma$ is either real or pure phase (see details in
next subsection) for given
$k$-values.
The obtained energy band structure for $N_z=9$ is shown in Fig.~\ref{figure:F2}(b).

For convenience, let us define several energy scales. \\
(i) $\varepsilon_{max}$: The top of the highest conduction band. This
also gives the bottom of the lowest valence band as $-\varepsilon_{max}$. 
For graphene, $\varepsilon_{max}=3$. For finite $N_z$, we have~\cite{note2}
$\varepsilon^2_{max}\approx 
5+4\cos\left(\frac{3\pi}{3N_z+2}\right)$. For large $N_z$, we obtain $\varepsilon_{max}\approx 3
[1-(\frac{\pi}{3N_z})^2]$. As we can see from Fig.~\ref{figure:F2}(b),
there is no propagating mode, i.e. only evanescent modes, for $|\varepsilon|>\varepsilon_{max}$. \\
(ii) $\Delta$: The energy that gives the range of the single-channel
region, in which $|\varepsilon|\le\Delta$.\\
(iii) $\Delta_0$: The energy for the lowest conduction subband
at $k=0$. 
As can be seen from Fig.~\ref{figure:F2}(b), 
there are only propagating modes (in total $2N_z$: $N_z$ left-going and $N_z$
right-going modes), i.e. no evanescent mode, in 
the energy range of $1\le|\varepsilon|<\Delta_0$.

\subsection{Propagating and evanescent modes}
For a given energy $\varepsilon$, once we obtain $\Lambda$ using
Eqs.~(\ref{poly}) and (\ref{lam}), a pair of $\lambda$s are obtained through the following relation:
\begin{equation}
\lambda_{\pm} = \frac{1}{2}\left\{\Lambda^2-2 \pm\sqrt{\Lambda^2(\Lambda^2-4)}\right\}. 
\label{lambda}
\end{equation}
If $0<\Lambda^2<4$, $\lambda$ takes complex values with
$|\lambda|=1$, which gives rise to propagating modes in the $x$-direction. 
Otherwise, $|\lambda|\neq 1$ and evanescent modes will appear instead.
We shall use $\lambda(+)$ to denote right-going (propagating or
decaying to the right) modes and $\lambda(-)=\lambda^{-1}(+)$ for
left-going (propagating or decaying to the left) modes. Note that one value
of $\sigma$ is accompanied by a pair of BMs represented by $\lambda(\pm)$.  

 \begin{table*}
 \caption{\label{table:1} Classification of the roots of Eq.~(\ref{poly}): $\sigma = e^{i\phi-\eta}$; 'phase': $|\sigma|=1$; 'complex': $\mbox{Im}(\sigma)\neq 0$; 'N.S.C.': no special constraints. $\varepsilon_{max}$ is given in the main text. The first two columns express necessary and sufficient conditions, but the conditions in the last four columns may not be sufficient. We distinguish edge states from extended states, which both are propagating modes.}
 \begin{tabular}{c c c c c c c c}
\hline\hline
Type & $|\lambda|$ & $\Lambda^2$ & $\sigma$ & $\phi$ & $\eta$ & $|\varepsilon|$ \\ [0.5ex] 
\hline
Edge & $=1$ & $\in[0,1]$ & real & $0,\pi$ & N.S.C. & $<\frac{1}{N_z+1}$ \\   
Extended & $=1$ & $\in[0,4]$ & phase & $\in(-\pi,\pi)$ & $0$ & $<\varepsilon_{max}$\\
Evanescent & $\neq 1$ & $\notin[0,4]$ & complex & $\neq 0,\pi$& N.S.C. & $\notin[1,\Delta_0]$ \\ 
\hline
 \end{tabular}
 \end{table*}

Equation (\ref{poly}) contains $4N_z$ roots of $\sigma$ for a given energy
$\varepsilon$. However, we show that these roots are four-fold redundant and there are actually
only $N_z$ physically distinguishable roots, which obtain  
$2N_z$ BMs (counting both left- and right-going modes, i.e., $\lambda(\pm)$) 
as expected on general grounds.~\cite{ando1991quantum,khomyakov2005conductance} 
To this end, we observe that Eq.~(\ref{poly})
obeys two symmetries: invariance under $\sigma \rightarrow \frac{1}{\sigma}$ and under $\sigma\rightarrow -\sigma$. Therefore, if
$\sigma$ is a root, then $\frac{1}{\sigma}$ and $-\sigma$ as well as $-\frac{1}{\sigma}$ must also be roots. 
Nevertheless, the BMs represented by these roots are physically identical (up to an irrelevant phase factor). In fact, we have 
$\hat{\psi}(m,n;
 {\lambda_{\pm},\sigma^{-1},s})=-\hat{\psi}(m,n;{\lambda_{\pm},\sigma, s})$
 and $\hat{\psi}(m,n;
 {\lambda_{\pm},-\sigma,s})=(-1)^{N_z+1}\hat{\psi}(m,n;{\lambda_{\pm},\sigma, s})$,
as can be deduced from Eqs.~(\ref{finalpsi})-(\ref{Q}). 
Thus, in spite of that Eq.(\ref{poly}) contains $4N_z$ roots, 
we have only $N_z$ physically distinguishable roots (and hence $2N_z$ BMs) owing to these
symmetries. 

Below we analyze the roots of Eq.~(\ref{poly}) in detail.
The results are summarized in Table \ref{table:1}.   
In general, $\sigma$ is a complex value, i.e. $\sigma = e^{ip}$ with $p = \phi + i\eta$.
According to above analysis, $p$ and $-p$ as well as $p+\pi$ give the same mode. It then suffices to focus on the
domain where $\eta>0$ and $0\leq\phi<\pi$, which contains a complete
set of BMs for ZGRs. 
No multiple roots exist for Eq.~(\ref{poly}) at any
$\varepsilon\neq 0,1$.~\cite{note1}

When $\sigma$ is real, i.e., $\phi = 0$ (or equivalently $\pi$), then $0<\Lambda^2<1$,
always leading to propagating modes. Actually, 
$p=i\eta$ corresponds to the
edge states that exist at very low energies. Indeed, Eq.~(\ref{poly}) gives 
\begin{equation}
|\varepsilon| = \frac{e^{-\eta N_z}}{\mathcal{G}_{N_z}(e^{-2\eta})},
\end{equation}
The right-hand side of this equation is monotonically decreasing for $\eta \ge
0$, whose maximum appears at $\eta=0$ and its value is 
$\frac{1}{N_z+1}$. This means that the edge states exist only if $|\varepsilon| \leq \varepsilon_c$, where
\begin{equation}
\varepsilon_c=\frac{1}{N_z+1}.  
\end{equation}
The $|\Lambda|$ at $\varepsilon_c$ can be obtained 
as $|\Lambda| = \frac{N_z}{N_z+1}$ 
by taking the limit of $\eta \rightarrow 0$ for Eq. (\ref{lam}).
Using $\lambda = e^{ik}$ for propagating modes, 
the wavenumber corresponding to $\varepsilon_c$ is then given by
\begin{equation}
2\cos\left(\frac{k_c}{2}\right) = \frac{N_z}{N_z+1},
\end{equation}
according to Eq.~(\ref{lamdatoLambda}).
Thus, the condition of real value for $\sigma$ determines the region for the
edge states, i.e. $|k|\geq k_c$.
This result is consistent with previously reported results.~\cite{wakabayashi2012nanoscale}

Next we consider when $\sigma$ is a pure phase factor, i.e., $\eta = 0$.
In this case, $\Lambda = -\frac{\sin(N_z\phi)}{\sin[(N_z+1)\phi]}$ [as
inferred from Eqs.~(\ref{S}) and (\ref{lam})] is real and both
propagating and evanescent modes can appear depending on whether $|\Lambda|\le 2$ or not. Actually, for $\phi$ close to the
nodes of $\sin(N_z\phi)$, we find small $|\Lambda|$ and hence
propagating modes; whereas for $\phi$ close to the nodes of
$\sin[(N_z+1)\phi]$, large $|\Lambda|$ results and we get evanescent
modes. Within the range of $0\le \phi \le \pi$, there are $N_z-1$ solutions of $p=\phi$ satisfying
Eq.~(\ref{lam}) if $|\Lambda|<1$, whereas $N_z$
solutions if $|\Lambda|>1$. The missing solution corresponds to the
edge state discussed in previous paragraph.~\cite{wakabayashi2012nanoscale}

Another special limit is $\sigma\rightarrow 0$, in which $\Lambda\sim
-\sigma\rightarrow 0$. This then corresponds to $k\rightarrow\pi$, the
completely localized edge state, whose energy is exactly zero by 
Eq.~(\ref{E}). Now Eq.~(\ref{poly}) becomes simply
$\sigma^{2N_z}\rightarrow0$, which has $2N_z$ identical roots
$\sigma\rightarrow 0$. 
The only BM is the completely localized edge state, whose wave function vanishes everywhere except on the zigzag
edges. Thus, the interior of the ZGR becomes completely irrelevant. We
expect this picture to be reasonable even for small but non-vanishing
$\varepsilon$. This observation has been recently utilized to account
for a parity effect~\cite{akhmerov2008,wakabayashi2002intb} occuring in ZGR p-n junctions.~\cite{deng2013jpsj}

Since propagating modes carry the flux of current, they are directly
 involved in scattering and transport problems. In band structure, these modes are usually labeled by wave
numbers $k$ together with $p$ and $s$, namely, 
\begin{equation}
\hat{\psi}({m,n;k,p,s}):=e^{ikm}\hat{\Phi}({n;k,p,s}),
\label{propsi}
\end{equation}
and the corresponding energy is denoted by 
$\varepsilon_{ps}(k)$. The group velocity is given as
\begin{equation}
v_{ps}(k) = \frac{\partial}{\partial k}\varepsilon_{ps}(k).
\label{v}
\end{equation} 
Right-going (left-going) propagating modes therefore have $v_{ps}(k)>0$
($v_{ps}(k)<0$). In Appendix B, we give a different expression for
$v_{ps}(k)$, which is more useful in numerical computations.

\subsection{Numerical analysis for $\sigma$}
Figures~\ref{figure:F2} (c)-(e) show the distribution of roots of
Eq.~(\ref{poly}) in the complex $\sigma$-plane for some specific
energies. Here we have numerically 
evaluated the roots of Eq.~(\ref{poly}) using the Durant-Kerner-Aberth
method.~\cite{kerner1966gesamtschrittverfahren,durand1972solutions}  The corresponding $\Lambda^2$ are calculated
according to Eq.~(\ref{lam}) to characterize the nature (propagating or
evanescent) of the roots. The behaviors of the roots depend on
the energy region.

\begin{enumerate}
 \item 
For $0<|\varepsilon|< \varepsilon_c$, none of the roots sit on the
   unit circle, see Fig.~\ref{figure:F2}(c). 
The elevation of energy (along the arrows) shifts the roots toward the circle. 
In the plot, the energies are taken as $\varepsilon=0.01,0.04,0.07,0.10$ for the case of $N_z=9$. 
Note that $\varepsilon=0.10\equiv \varepsilon_c$ for $N_z=9$. 

\item
In the range $\varepsilon_c\le |\varepsilon|< 1$, only those roots which
represent propagating modes are located on the unit circle, see
     Fig.~\ref{figure:F2}(d). 
In this plot, the energies are taken as $\varepsilon=0.1,0.4,0.7,1.0$.
The arrow indicates the direction of increasing energy.

\item
In the region of $1\le \varepsilon \le \Delta_0$, 
all roots give rise to propagating modes and they lie on the unit circle. 
In the region of $|\varepsilon|\ge \Delta_0$, the roots still sit
on the circle but their phases get shifted and some of them are converted into evanescent modes, see Fig.~\ref{figure:F2}(e). 
In this plot, the energies are taken as $\varepsilon=1.0,1.5,2.0,2.5,3.0$.
\end{enumerate} 

\subsection{Analytical expressions for low-energy roots}
The foregoing analysis renders analytical expressions for all the roots
at $|\varepsilon|\ll 1$. To see this, we at first look at
$|\varepsilon|\leq \varepsilon_c$. In such case, all roots have
$|\sigma|\neq 1$ and they come in pairs, $(\sigma,\sigma^{-1})$. It
suffices to find out those lying inside the unit circle, i.e.,
$|\sigma|\leq 1$. We denote such roots by $x_r$, where
$r=0,...,N_z-1$. Since $|x_r|<1$, the expression $\mathcal{G}_{N_z}(x)$ in
Eq.~(\ref{poly}) may be approximated as $\mathcal{G}_{N_z}(x)\approx 1$ and
hence, we obtain $x^{N_z}_r\approx |\varepsilon|$. From this, we find  
\begin{equation}
x_r \approx |\varepsilon|^{\frac{1}{N_z}}e^{i\frac{\pi}{N_z}r}, 
\label{x}
\end{equation}
and then the complete set of roots is given as
\begin{equation}
\{x_0,x^{-1}_0,x_1,x^{-1}_1,...,x_{N_z-1},x^{-1}_{N_z-1}\}, \quad\mbox{for $|\varepsilon|<\varepsilon_c$}.
\label{small}
\end{equation}
The pair that gives the edge state is $(x_0,x^{-1}_0)$. 

By increasing $|\varepsilon|$ above $\varepsilon_c$ but below $\Delta$,
the edge state becomes an 
extended state, i.e., the pair $(x_0,x^{-1}_0)$ is displaced onto the
unit circle: $(x_0,x^{-1}_0)\rightarrow
(e^{i\phi_0},e^{-i\phi_0})$. Here $\phi_0$ is the phase angle to be
worked out later. On other hand, all other roots can still be well
described by $(x_r,x^{-1}_r)$, as is clear in Fig.~\ref{figure:F2}(d). Thus,
we find the complete set of roots to be 
\begin{equation}
\{e^{i\phi_0},e^{-i\phi_0},x_1,x^{-1}_1,...,x_{N_z-1},x^{-1}_{N_z-1}\},
 \mbox{for $|\varepsilon|\in[\varepsilon_c,\Delta)$}. 
\label{big}
\end{equation} 
To obtain $\phi_0$, we substitute $\sigma=e^{i\phi_0}$ in Eq.~(\ref{poly}) and find
\begin{equation}
\left|\frac{\sin\phi_0}{\sin (N_z+1)\phi_0}\right|=|\varepsilon|,
\label{phi0}
\end{equation} 
whose solution always lies in $(0,\frac{\pi}{N_z+1})$ for any
$|\varepsilon|\in(\varepsilon_c,\Delta)$. For $|\varepsilon|$ slightly
above $\varepsilon_c$, $\phi_0$ is very small and we can then expand the
left-hand side of this equation to get
\begin{equation}
\phi_0\approx \frac{1}{N_z+1}\sqrt{\frac{6(|\varepsilon|/\varepsilon_c-1)}{|\varepsilon|/\varepsilon_c-(N_z+1)^{-2}}}.
\label{phi0app}
\end{equation}
Although this expression has been derived by assuming that
$|\varepsilon|$ be close to $\varepsilon_c$, it can actually describe
the solution very accurately in the whole regime
$(\varepsilon_c,\Delta)$. In the limit $N_z\rightarrow\infty$, we find
$\phi_0\rightarrow \frac{\sqrt{6}}{N_z+1}$ for fixed $|\varepsilon|$,
very close to the exact value of $\frac{\pi}{N_z+1}$. 

By analogy with Eq.~(\ref{big}), one can write down sets of roots for
even higher energy, but analytical expressions for the roots (more than
one) on unit circles [i.e., of Eq.~(\ref{phi0})] become impossible. The
results contained in (\ref{small}) and (\ref{big}), supplemented by
Eqs.~(\ref{x}) and (\ref{phi0}), provide a complete foundation for
studying e.g. quantum transport in a variety of ZGR junctions in the
single-channel regime. There is no need to numerically search for the
roots of Eq.~(\ref{poly}) any more. The interrelation [Eq.~(\ref{lamsig})]
between $\sigma$ and $\lambda$ becomes effectively dissolved.   

\section{Applications}
\label{section:4}
This section is devoted to a few examples which illustrate common
applications of the BMs we derived so far for graphene and graphene
ribbons. A straightforward example is shown in the following
subsection, where the electronic structure of bearded ZGRs is
derived. Bearded ZGRs~\cite{klein1994graphitic,wakabayashi2001prb} are
ZGRs with extra sites attached to one of its zigzag edges and they may
be realized by chemical modifications. In another example presented in
the second subsection we study quantum transport through a graphene
aperture~\cite{deng2014prb} using mode-matching method.~\cite{ando1991quantum} The BMs are then directly used in
obtaining the transmission matrix. With the analytical prescriptions of
the BMs, we can easily simulate ZGRs of width of hundreds of zigzag
chains. Finally, in the third example considered in the third
subsection, we relate the BMs to the core quantities of the widely
employed nonequilibrium Green's function method~\cite{datta1997electronic} in transport
simulations. We explicitly construct the Green's functions in terms of BMs.  

\subsection{Electronic structure of singly bearded ZGRs}
\begin{figure}
\includegraphics[width=0.48\textwidth]{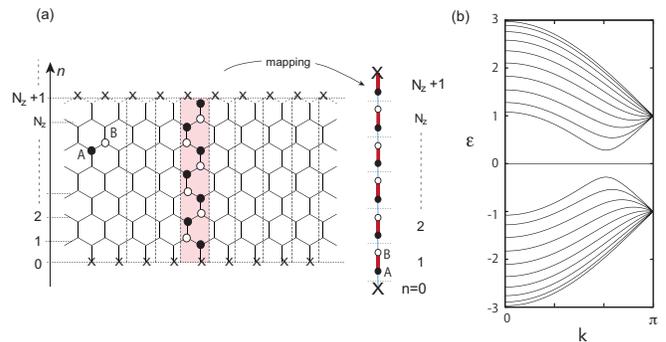}
\caption{(a) The lattice structure of singly bearded ZGR.
(b) Energy band structure for $N_z=10$. The central flat subband extends throughout the 1BZ.
}
\label{figure:F3}
\end{figure}

Here we shall show that our approach correctly
describes the energy spectrum and wave functions of singly bearded ZGRs. 
The lattice structure and energy band structure are shown in
Figs.~\ref{figure:F3} (a) and (b), respectively. 
Since the system has the translational invariance along $x$-direction, 
we may assume that $\lambda = e^{ik}$. 
The boundary conditions requires for the wave functions can be written as 
$\psi_B(m,0)$ and $\psi_B(m,N_z+1)$. 

For $\varepsilon\neq 0$, the $\sigma$ can be shown to be a phase factor
due to the boundary conditions. Thus, we put $\sigma=e^{ip}$. In analogy with
Eq.~(\ref{psi}), we can easily derive the wave functions to be 
\begin{equation}
\hat{\psi}(m,n;k,p,s)\propto e^{ikm} 
\hat{M} 
\begin{pmatrix}
s\cdot\sin(pn-\theta) \\
\sin(pn)
\end{pmatrix},
\label{beard1}
\end{equation}
where $\theta$ is the phase angle of $\gamma(k,p)\eqqcolon
g_k+e^{ip}$ and $p = \frac{\pi}{N_z+1}\cdot r$, with
$r=1,2...,N_z$. The energy of this state is given by $\varepsilon_{kps}
= s|\gamma(k,p)|$, by Eq.~(\ref{E}). Note that $p$ does not depend on
$k$, differing from the case with ZGR and doubly bearded ZGR.~\cite{klein1994graphitic} 

For $\varepsilon =0$, we have $\sigma = - g_k$ according to
Eq.~(\ref{E}). The possibility that $\sigma =
- g^{-1}_k$ is excluded by the boundary conditions. Denote the zero
state by $\hat{\psi}_0(m,n)$ and we find 
\begin{equation}
\hat{\psi}_0(m,n) \propto e^{ikm}\hat{M}
\begin{pmatrix}
(-2\cos\frac{k}{2})^n \\
0
\end{pmatrix}.
\label{beard2}
\end{equation}
Here $k$ can take any value between $-\pi$ and $\pi$, because
$n=1,2,...,N_z+1$ is finite. Expectedly, these states are localized
about the bearded edge if $|k|<\frac{2\pi}{3}$ but on the unbearded edge
if $|k|>\frac{2\pi}{3}$.   

Figure~\ref{figure:F3}(b) shows the energy band structure for 
$N_z=10$.~\cite{wakabayashi2001prb}
The central subband is completely flat, separated
from the adjacent subbands by a gap of $\sin(\frac{\pi}{N_z+1})$ that is
reached near the Dirac point.

\subsection{Resonant transport through an aperture}
\begin{figure*}
\includegraphics[width=\textwidth]{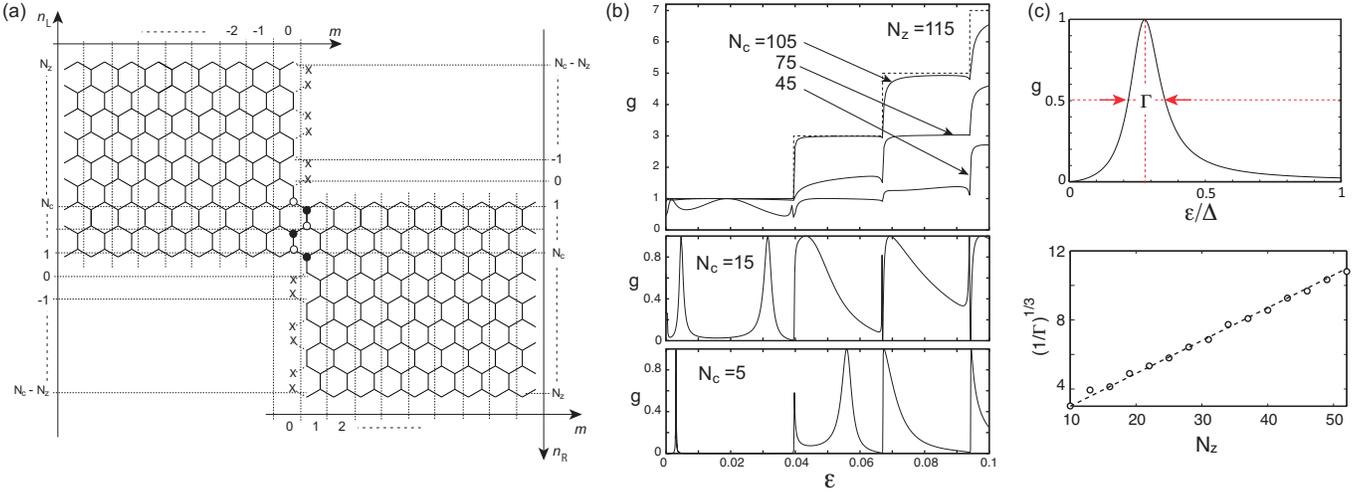}
\caption{(a) Structure of a graphene aperture connecting two graphene 
nanoribbons (width $N_z$), where the aperture consists of $N_c$
 connecting bonds.
 We use $n_L$ and $n_R$ to label the
 zigzag chains for the left and right nanoribbons, respectively. 
 The local wave functions vanish on crossed sites while they match on the sites on
 connecting bonds marked by black and white circles.
(b) The energy dependence of dimensionless conductance $g$ for
 $N_z=115$ and various $N_c$. 
(c) For $N_c\ll N_z$, the $g$ shows a
 resonance of width $\Gamma$ that scales as $N^{-3}_z$. The $\Delta$ is
 indicated in Fig.~\ref{figure:F2} (b).} 
\label{figure:F4}
\end{figure*}

In this subsection, we employ the BMs of ZGR to study electronic
transport through a graphene aperture [see Fig.~\ref{figure:F4}
(a)]. This type of aperture was studied by us~\cite{deng2014prb} and it
was analytically shown that peculiar low-energy resonant states could
form due to the presence of edge states. In what follows, we revisit
this phenomenon by a mode-matching approach. We perform  numerical
computations, on the basis of Eq.~(\ref{poly}), to obtain the
conductance through the aperture. 

As clear from the energy band structure shown in Fig.~\ref{figure:F2}(b), at a
given $\varepsilon$, a propagating mode is specified by its
$k$, which takes on discrete values
$k_1,\bar{k}_1,...,k_{N_p},\bar{k}_{N_p}$. 
Here we reserve $k_i$ for the
$i$-th right-going mode while $\bar{k}_i=-k_i$ for the coresponding
left-going mode. 
In general, we have $2N_p$ propagating modes, where 
factor $2$ arises due to the fact that left-going and right-going modes
come together.
Accordingly, there are $N_e = N_z - N_p$ pairs of evanescent modes.  
In particular, for $|\varepsilon|<\Delta$, we have $N_p=1$. 

The size of the transmission matrix
$\hat{t}(\varepsilon)$ is $N_p\times N_p$. 
Thus, let us denote the transmission coefficient from $k_i$ to $k_j$ mode as
$t_{ji}(\varepsilon):= t_{k_j,k_i}(\varepsilon)$, where
$i,j=1,...,N_p$. 
The conductance $G$ is given by the Landauer-B\"uttiker formula:
\begin{equation}
G = \frac{2e^2}{h}\sum_{j=1}^{N_p}\sum_{i=1}^{N_p}|t_{ji}|^2 \eqqcolon \frac{2e^2}{h} g,
\end{equation}
where $g$ is the dimensionless conductance.

Suppose an electron is injected upon the aperture from left at energy
$\varepsilon>0$ and with wave number $k_i$. It will be partially
reflected into a mode of wave number $\bar{k}_{i^\prime}$ with amplitude
$r_{i^\prime i}$ and partially transmitted into a mode of wave number $k_j$
with amplitude $\tilde{t}_{ji}$. Simultaneously, evanescent modes decaying away
from the aperture will be excited. On the left-hand side to the
aperture, the wave function can then be written as 
\begin{eqnarray}
\hat{\Psi}^L({m,n_L;k_i}) &=& e^{ik_im}\hat{\Phi}({n_L;k_i}) \nonumber\\
&+&\sum^{N_p}_{i'=1}r_{i^\prime i}e^{i\bar{k}_{i'}m}\hat{\Phi}({n_L;\bar{k}_{i'}}) 
+ \sum^{N_e}_{I=1}C^L_{Ii}\lambda^m_I\hat{\Phi}({n_L;I}),\nonumber\\
\label{PsiL}
\end{eqnarray}
where $\hat{\Phi}({n;k_i}):=\hat{\Phi}({n;k_i,p_i,+})$ [see
Eq.~(\ref{propsi})],
$\hat{\Phi}({n;I}):=\hat{\Phi}({n;\lambda_I,\sigma_I,+})$ and $C^L_{Ii}$
denotes the amplitudes for the $I$-th evanescent mode. Moreover, we
require $v_{p_i+}(k_i)>0$ and $v_{p_i+}(\bar{k}_i)<0$ as well as
$|\lambda_I|>1$. Similarly, on the right-hand side, the wave function is
given by 
\begin{equation}
\hat{\Psi}^R({m,n_R;k_i}) =
 \sum^{N_p}_{j=1}\tilde{t}_{ji}e^{ik_jm}\hat{\Phi}({n_R;k_j}) +
 \sum^{N_e}_{J=1}C^R_{Ji}\lambda^m_{J}\hat{\Phi}({n_R;J}). 
\label{PsiR}
\end{equation}
Here we must have $|\lambda_J|<1$. The $n_{L/R}$ in Eqs.~(\ref{PsiL})
and (\ref{PsiR}) are defined in Fig.~\ref{figure:F4} (a).
The $\tilde{t}_{ij}$ can be related to $t_{ij}$ as follows, 
\begin{equation}
t_{ji} = \sqrt{\frac{v_{p_j+}(k_j)}{v_{p_i+}(k_i)}} \cdot \tilde{t}_{ji}. 
\label{T}
\end{equation}
It should be noted that if the number of total connecting bonds ($N_c$) is even, the matrix $\hat{M}$ defined via Eq.~(\ref{M}) has to be replaced by its inverse in calculating $\hat{\Psi}^R({m,n_R;k_i})$.

The amplitudes $r_{i'i}$ and $\tilde{t}_{ji}$ are uniquely determined by
the matching conditions appropriate for the aperture shown in
Fig.~\ref{figure:F4} (a). Firstly, the $\hat{\Psi}^L({m,n_L;k_i})$ must
vanish on all the crossed sites lying on $m=1$, i.e., 
\begin{eqnarray}
\Psi^L_A\left({1,N_c+2r+mod(N_c,2)+1;k_i}\right)&=&0,\nonumber\\
\Psi^L_B\left({1,N_c+2(r+1)-mod(N_c,2);k_i}\right)&=&0.
\label{van1}
\end{eqnarray}
Besides, the $\hat{\Psi}^R({m,n_R;k_i})$ must vanish on all crossed sites lying on $m=0$. Thus,
\begin{eqnarray}
\Psi^R_A\left({0,N_c+2r+1;k_i}\right)&=&0,\nonumber \\
\Psi^R_B\left({0,N_c+2(r+1);k_i}\right)&=&0.
\label{van2}
\end{eqnarray}
In the above, $r=0,...,[N_z-N_c-mod(N_z-N_c,2)]/2-1$. Secondly, the $\hat{\Psi}^L({m,n_L;k_i})$ and $\hat{\Psi}^R({m,n_R;k_i})$ must be equal to each other on all the sites linked by the connecting bonds. Namely, we have
\begin{eqnarray}
\Psi^L_A\left({0,2(w+1);k_i}\right)&=&\Psi^R_A\left({0,N_c+1-2(w+1);k_i}\right),\nonumber\\
\Psi^L_B\left({0,2w+1;k_i}\right)&=&\Psi^R_A\left({0,N_c-2w;k_i}\right),\nonumber\\
\Psi^L_A\left({1,2w+1;k_i}\right)&=&\Psi^R_A\left({1,N_c-2w;k_i}\right),\nonumber\\
\Psi^L_B\left({1,2(w+1);k_i}\right)&=&\Psi^R_B\left({1,N_c+1-2(w+1);k_i}\right),
\label{van3}
\end{eqnarray}
where $w=0,...,[N_c-mod(N_c,2)]/2-1$. Altogether, Eqs.~(\ref{van1}) - (\ref{van3}) produce $2N_z$ independent 
conditions, which uniquely determine the $2N_z$ unknowns, $r_{i'i}$,
$\tilde{t}_{ji}$, $C^L_{Ii}$ and $C^R_{Ji}$.~\cite{deng2013jpsj} 

We have numerically calculated the Landauer conductance for a variety of
structural parameters for the aperture. Since we know the analytic form
of BMs, we can easily reach the relatively larger system over
$N_z=100$ with very little computation time. 
Fig.~\ref{figure:F4}(b) shows the energy dependence of dimensionless
conductance $g$ for $N_z=115$ with various $N_c$. The decrease of $N_c$
causes the reduction of conductance. In the region of $N_c\ll N_z$, the
resonances appear in the low energy region. The origin of these
resonances is attributed to the formation of quasi-bound states near the
aperture due to the interference of edge states.~\cite{deng2014prb}
A typical behavior of lowest-energy resonance for $N_c\ll N_z$ is
displayed in Fig.~\ref{figure:F4}(c)
Its width at the maximum is scaled 
with $N_z$ following a power law, $\Gamma^{-1} \sim N^3_z$, in agreement
with our analytical treatment.~\cite{deng2014prb}

\subsection{Green's functions of ZGR electrodes}
In preceding subsection, the utility of BMs has been elucidated
with the mode-matching method~\cite{ando1991quantum} and the conductance
$G$ of a graphene aperture was computed. In the literature, however,
this conductance is more frequently evaluated by an apparently different
method based on Green's functions.~\cite{datta1997electronic} The
equivalence between the mode-matching and the Green's function methods
have been demonstrated in Ref.\cite{khomyakov2005conductance}. In what
follows we explain how to obtain the Green's functions in terms of the
ZGR BMs. 

\begin{figure}
\includegraphics[width=0.48\textwidth]{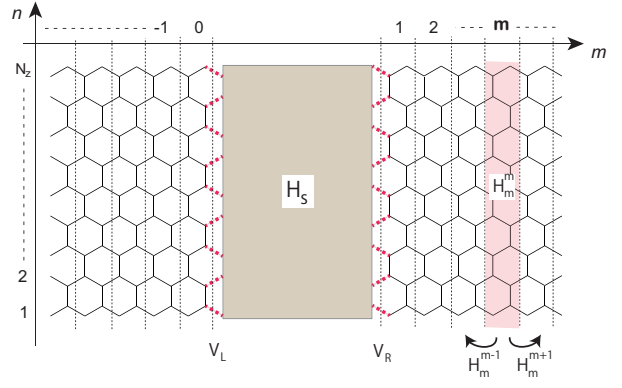}
\caption{A generic setup in two-probe transport measurements. The conducting channel, represented by the Hamiltonian $H_s$, is connected to the left electrode and the right electrode via couplings $V_L$ and $V_R$, respectively. Both electrodes are represented by semi-infinite ideal ZGRs each consisting of $N_z$ zigzag chains.}
\label{figure:F5}
\end{figure}

We consider a general two-probe experimental setup as portrayed in
Fig.~\ref{figure:F5}, where a conducting channel or scattering region
(represented by $H_s$) is connected to two electrodes via $V_L$ and
$V_R$, respectively. Both electrodes are represented by semi-infinite
ZGRs. The left electrode extends from $m=0$ to $m=-\infty$, whereas the
right one from $m=1$ to $m=\infty$. The $V_L$ ($V_R$) couples only the
supercell $m=0$ ($m=1$) to the channel. We shall write the Hamiltonian
for the left electrode as $H_L = \sum^{-\infty}_{m=0}(H^m_m\oplus
H^{m-1}_{m})$. Similarly, for the right electrode,
$H_R=\sum^{\infty}_{m=1}(H^m_m\oplus H^{m}_{m+1})$. Here $H^m_m$ couples
sites in the same supercell, while $H^m_{m+1}$ couples sites in the
$m$-th supercell to those in the $(m+1)$-th one (see
Fig.~\ref{figure:F5}). Since the block matrice $H^m_{m\pm 1}$ are
independent of $m$, we then put $V\equiv H^m_{m-1}$ and
$V^{\dagger}\equiv H^{m}_{m+1}$. Further, we define a global block
Green's function as $G_{m,m'}(\omega)=\langle
m|(\omega-H_s-H_L-H_R)^{-1}|m'\rangle$, where $\omega$ is a complex
parameter and $H_s$ has included both $V_L$ and $V_R$ by definition. Now
the conductance of the setup can be evaluated as
\cite{datta1997electronic,caroli1971direct}
\begin{equation}
g = \frac{2e^2}{h} Tr[\Gamma_RG^r\Gamma_LG^a],
\label{G}
\end{equation}  
where $G^{r} = G_{1,0}(\varepsilon+i0_+)$, $G^a = G_{0,1}(\varepsilon-i0_+)$. In addition, the broadening functions $\Gamma_{R/L}$ are given by
\begin{equation}
\Gamma_{R/L} = i(\Sigma_{R/L}-\Sigma^{\dagger}_{R/L}).
\end{equation}
Below we prescribe the expressions for the self-energies $\Sigma_{R/L}$
in terms of the BMs.

At given energy $\varepsilon$, $s$ is fixed and there are in total $N_z$ values of $\sigma$, i.e., $\sigma_1,...,\sigma_{N_z}$. For each $\sigma_u$, where $u=1,...,N_z$, there is a pair of $\lambda$s [see discussions following Eq.(\ref{lambda})], $\lambda_u(+)$ and $\lambda_u(-)=\lambda^{-1}(+)$ standing for right-going and left-going modes, respectively. 
We define a number of column vectors by Eq.~(\ref{Phi}),
\begin{equation}
 \Phi_{u}(\pm):=[\Phi_A({1;u}),\Phi_B({1;u}),...,\Phi_A({N_z;u}),\Phi_B({N_z;u})]^T,
\end{equation}
where $\Phi_{\nu}({n;u}):=\Phi_{\nu}({n;\lambda_u(\pm),\sigma_u,s})$. Further,
we introduce $N_z$ pairs of dual vectors, $\tilde{\Phi}_u(\pm)$,
uniquely defined by 
\begin{equation}
\tilde{\Phi}^{\dagger}_u(\pm)\Phi_{u'}(\pm) = \delta_{uu'},\quad \Phi^{\dagger}_u(\pm)\tilde{\Phi}_{u'}(\pm)=\delta_{uu'}.
\label{dual}
\end{equation} 
Finally, we need the following $2N_z\times 2N_z$ matrix,
\begin{equation}
\mathcal{K}^r(\pm) = \sum^{N_z}_{u=1} \lambda^r_u(\pm) \Phi_u(\pm) \tilde{\Phi}^{\dagger}_u(\pm),
\end{equation}
with $r$ being an integer. For completeness, we sketch how to find the
$\tilde{\Phi}_u(\pm)$ from $\Phi_u\equiv [\Phi_u(+),\Phi_u(-)]$. Let us
define a $2N_z\times 2N_z$ matrix, $\mathcal{L}$, whose elements are
given by $\mathcal{L}^i_{j} = \Phi^{\dagger}_j(i)$, and a number of
$2N_z\times 1$ column vectors, $D_{j}(\pm)$, whose elements are given by
$D_j(+,i\leq N_z)=\delta_{ij}$, $D_j(+,i>N_z)=0$, $D_j(-,i\leq N_z)=0$
and $D_j(-,i>N_z)=\delta_{i-N_z,j}$. Then, from Eq.~(\ref{dual}), we get
$\tilde{\Phi}_u(\pm) = \mathcal{L}^{-1} D_u(\pm)$. 

With the above definitions, the self-energies can then be obtained as,~\cite{khomyakov2005conductance}
\begin{equation}
\Sigma_L = V\mathcal{K}^{-1}(-), \quad \Sigma_R = V^{\dagger}\mathcal{K}^1(+). 
\label{selfenergy}
\end{equation}
We remark that the $G^{r/a}$ in Eq.~(\ref{G}) can be simplified by using these self-energies~\cite{datta1997electronic}. Additionally, the block Green's function for the left semi-infinite ZGR can be written as
\begin{eqnarray}
G^L_{m,0}(\varepsilon) &\equiv& \langle m|\frac{1}{\varepsilon+i0_+-H_L}|0\rangle\nonumber\\ &=& \mathcal{K}^m(-) g(\varepsilon+i0_+), \quad m<0,
\label{GL}
\end{eqnarray}  
with $g(\varepsilon+i0_+)$ being the surface Green's function given by
\begin{equation}
g(\varepsilon+i0_+) = \mathcal{K}^{-1}(-)(V^{\dagger})^{-1}.
\label{surfaceG}
\end{equation}
Analogously, for the right semi-infinite ZGR, we have
\begin{equation}
G^R_{m,1}(\varepsilon)\equiv\langle m|\frac{1}{\varepsilon+i0_+-H_R}|1\rangle = \mathcal{K}^{m}(+)V^{-1}, \quad m\geq 1. \label{GR}
\end{equation}
Moreover, the retarded Green's function for an infinite ideal ZGR, $G^0_{m,m'}(\varepsilon)$, can also be easily obtained using the bulk modes. We have~\cite{khomyakov2005conductance}
\begin{eqnarray}
G^0_{m,m'}(\varepsilon) &=& \{\theta(m-m')\mathcal{K}^{m-m'}(+)+\theta(m'-m)\mathcal{K}^{m-m'}(-)\}\nonumber\\
&\cdot&\left\{V^{\dagger}[\mathcal{K}(-)-\mathcal{K}(+)]\right\}^{-1},
\label{G0}
\end{eqnarray}
where $\theta(x)$ denotes the Heaviside step function. 

\section{Summary}
\label{section:5}
We have obtained complete sets of BMs for graphene and graphene ribbons
(both ZGR and AGR) at arbitrary energy $\varepsilon$. These modes are
presented in a representation that is particularly suitable for studying
the electronic conduction of graphene nanostructures. Their properties
have been thoroughly analyzed and analytical expressions for low-energy
ZGR BMs have been prescribed. At fixed energy, each BM can be specified
by a single parameter $\sigma$ (while the $\lambda$ is determined from
$\sigma$ and therefore not a free parameter), which (in ZGR) is shown to
satisfy a simple polynomial equation that can be efficiently solved with
existing numerical techniques. 

The utility of the BMs is illustrated in a few examples, by which we
demonstrated that they could be used to construct the electronic band
structure of graphene nanostructures, to calculate the transmission
matrix of graphene devices and to evaluate the Green's functions of both
infinite and semi-infinite graphene ribbons. These Green's functions
play an important role in modern mesoscopic transport theory. We also
note that the BMs can be used to construct the transfer matrix in the
so-called transfer-matrix
method,~\cite{ferreira2011,hernandez2012,grover2012} which is basically
a variation of the mode-matching method. 

Moreover, exploring the feature of the wave functions for BMs may offer
intuitive understanding of some otherwise perplexing phenomena, as
demonstrated in the case of transmission through bends and
polygons~\cite{iyengar2008} and current blocking effect in ZGR p-n
junctions.~\cite{deng2013jpsj}

\begin{acknowledgements}
K. W. acknowledges the financial support by Grant-in-Aid for Scientific
 Research from MEXT and JSPS (Nos. 25107001, 25107005 and 23310083).
\end{acknowledgements}

\appendix
\section{Bulk Modes and Green's functions of Armchair Graphene Ribbons}

\begin{figure}
\includegraphics[width=0.4\textwidth]{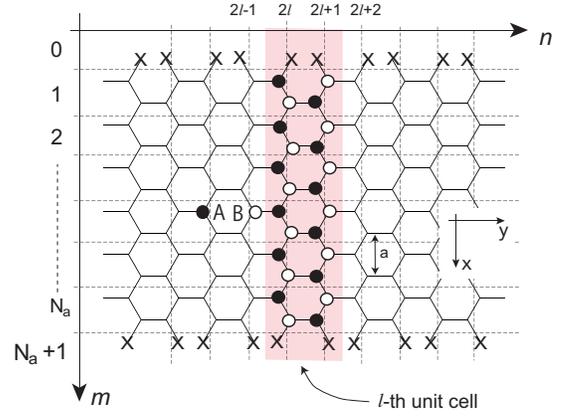}
\caption{Lattice structure of armchair graphene ribbons.}
\label{figure:F6}
\end{figure}

\subsection{Bulk Modes}
We consider an AGR extended in the $y$-direction but confined between
$m=0$ and $m=N_a+1$, as depicted in Fig.~\ref{figure:F6}.
We may include only the left half of the $(N_a+1)$-th supercell.
In close analogy with Eq.(\ref{psi}), we write the bulk
modes as 
\begin{equation}
\hat{\varphi}(m,n;{\lambda,\sigma,s}) \coloneqq \sigma^n \left(\beta \lambda^m
 \hat{M} - \beta' \lambda^{-m} \hat{M}^{-1} \right) \hat{f}_{\lambda\sigma s},
\end{equation} 
where $\beta$ and $\beta'$ are coefficients to be determined using the
boundary condition.
Here we have used that
$\hat{f}_{\lambda\sigma s}=\hat{f}_{\lambda^{-1}\sigma s}$ as well as
$M(\lambda^{-1}) = M^{-1}(\lambda)$. 
$\hat{\varphi}(m,n;{\lambda,\sigma,s})$ is a spinor,
which is defined as
\begin{equation}
\hat{\varphi}(m,n;{\lambda,\sigma,s}) = 
\left(
 \begin{array}{c}
\varphi_{\rm A}(m,n;{\lambda,\sigma,s}) \\
\varphi_{\rm B}(m,n;{\lambda,\sigma,s})
 \end{array}
\right).
\end{equation}
In the following, we simply write
$\hat{\varphi}(m,n)$ and ${\varphi}_\nu(m,n)$ ($\nu = {\rm A, B}$)
instead of 
$\hat{\varphi}(m,n;{\lambda,\sigma,s})$ and
${\varphi}_\nu(m,n;{\lambda,\sigma,s})$, respectively,  
if there is no ambiguity. 
The B.C. for AGR is written as
\begin{equation}
\begin{array}{ll}
\varphi_{\rm A}(0,2l-1)=0, &
\varphi_{\rm B}(0,2l  )=0, \\
\varphi_{\rm A}(N_a+1,2l-1)=0, &
\varphi_{\rm B}(N_a+1,2l)=0, 
\end{array}
\end{equation}
where $l$ is any integer (see Fig.~\ref{figure:F6}). 
Note that if the left half of the $(N_a+1)$-th supercell is
excluded, the latter two conditions have to be changed to
\begin{equation}
\begin{array}{ll}
\varphi_{\rm B}(N_a+1,2l-1)=0, &  \varphi_{\rm A}(N_a+1,2l)=0. 
\end{array}
\end{equation}
Nevertheless, basically the same results occur and
hence we here ignore this case. From these conditions, we find 
\begin{eqnarray}
\beta \lambda^{1/4} &-& \beta'\lambda^{-1/4} = 0,\nonumber \\
\beta \lambda^{1/4} \lambda^{N_a+1} &-& \beta'\lambda^{-1/4} \lambda^{-(N_a+1)}= 0.
\end{eqnarray} 
It immediately follows that 
\begin{equation}
\lambda^{2(N_a+1)} = 1,
\end{equation}
which demands $\lambda$ to be a simple phase factor. Thus, we may put $\lambda = \lambda_j = e^{ik_j}$, with $k_j = \frac{\pi}{N_a+1}\cdot j$, where $j=1,...,N_a+1$, leading to $(N_a+1)$ pairs of subbands. From this, the $\sigma = \sigma_j$ is then determined by Eq.(\ref{E}), which yields
\begin{equation}
\sigma_j \coloneqq \frac{1}{2}\left\{\zeta^2_j-2\pm\sqrt{\zeta^2_j(\zeta^2_j-4)}\right\},
\end{equation}
where $\zeta_j = \frac{\varepsilon^2-1-g^2_{k_j}}{g_{k_j}}$ with $g_k =
2\cos(k/2)$. We shall use $\sigma_j(+)$ [$\sigma_j(-) =
\sigma^{-1}_j(+)$] to denote the left- (right-) going modes.
Now we obtain 
\begin{eqnarray}
\label{varphi}
&\quad&
\hat{\varphi}_{js}(m,n) \coloneqq \hat{\varphi}(m,n;{\lambda_j,\sigma_j,s}) = \sigma^n_j \nonumber\\ 
&\times&
\begin{pmatrix}
\sin[k_j(m-\frac{(-1)^n+1}{4})] & 0 \\
0 & \sin[k_j(m+\frac{(-1)^n-1}{4})] 
\end{pmatrix}
\hat{f}_{\lambda_j\sigma_j s}\nonumber\\
\end{eqnarray}
where an irrelevant factor has been absorbed in the normalization factor of $f_{\lambda_j\sigma_j s}$. 

\subsection{Green's functions}
We proceed to establish the Green's functions of AGR by the BMs. To
this end, we need to adapt Eq.(\ref{varphi}) into a form that fits the
AGR unit cell as defined by the underlying translational symmetry. An
AGR unit cell, indexed by $l$, consists of two adjacent zigzag chains,
$n=2l$ and $n=2l+1$, instead of one. Let us consider the unit cell $l=0$, which contains 
the $0$-th and $1$-st zigzag chains. We may arrange the values of $\hat{\varphi}_{js}(m,0)$ into a $2(N_a+1)\time 1$ column vector $\Phi^0_j$, defined as 
\begin{eqnarray}
\Phi^0_j&:=&[\varphi_A({1,0;j}),\varphi_B({1,0;j}),...,\varphi_A({N_a,0;j}),\varphi_B({N_a,0;j}),\nonumber\\
&\quad&\varphi_A({N_a+1,0;j}),0]^T,
\end{eqnarray}
where $\varphi_{\nu}({m,n;j}):=\varphi_{\nu}({m,n;\lambda_j,\sigma_j,s})$. Similarly, we define
\begin{eqnarray}
\Phi^1_j&:=&[\varphi_A({1,1;j}),\varphi_B({1,1;j}),...,\varphi_A({N_a,1;j}),\varphi_B({N_a,1;j}),\nonumber\\
&\quad&\varphi_A({N_a+1,1;j}),0]^T.
\end{eqnarray} 
Now we merge $\Phi^0_j$ and $\Phi^1_j$ in a single $4(N_a+1)\times 1$ column vector,
\begin{equation}
\Phi_j := 
\begin{pmatrix}
\Phi^0_j \\
\Phi^1_j
\end{pmatrix},
\end{equation}
which is no more than the wave function of the $j$-th mode for the unit cell $l=0$. For an arbitrary unit cell, the wave function can also be written as a $4(N_a+1)\times1$ column vector, $\varphi_j(l)$, which isrelated to $\varphi_j(l=0)=\Phi_j$ as follows,
\begin{equation}
\varphi_j(l) = \tilde{\sigma}^l_j \Phi_j,
\end{equation}
where $\tilde{\sigma}_j = \sigma^2_j$. 
Following scrupulously Ref.[34], we hereafter
write $\Phi_j(\pm)$ in regard to
$\tilde{\sigma}_j(\pm)=\sigma^2_{j}(\pm)$. Now we define the
vectors $\tilde{\Phi}_j(\pm)$ which are dual to $\Phi_j(\pm)$, 
\begin{equation}
\tilde{\Phi}^{\dagger}_j(\pm)\Phi_{j'}(\pm) = \delta_{j,j'},\quad \Phi^{\dagger}_j(\pm)\tilde{\Phi}_{j'}(\pm) = \delta_{j,j'}.
\end{equation}
These $\tilde{\Phi}_j(\pm)$ can be written as a merger of two $2(N_a+1)\times 1$ vectors, $[\tilde{\Phi}^>_j(\pm),\tilde{\Phi}^<_j(\pm)]^T$, where $\tilde{\Phi}^<_j$ can be determined from $\tilde{\Phi}^>_j$, due to the fact that $\Phi_j(i)$ and $\Phi_j(i+2(N_a+1))$ are related, as is clear from Eq.~(\ref{varphi}). Finally, the vectors $\Phi_j(\pm)$ and $\tilde{\Phi}_j(\pm)$ are used to construct the $\mathcal{K}$ matrice,
\begin{equation}
\mathcal{K}^r(\pm) = \sum^{2(N_a+1)}_{j=1}\tilde{\sigma}^r_j(\pm) \Phi_j(\pm)\tilde{\Phi}^{\dagger}_j(\pm),
\label{mathK}
\end{equation}
with $r$ being an integer. Finally, with Eq.~(\ref{mathK}), the Green's
functions for infinite and semi-infinite AGRs can be written down in
exactly the same form as Eqs.~(\ref{selfenergy})-(\ref{G0}), except that one replaces $m$ by
$l$ and $V$ by the coupling between adjacent AGR unit cells.   

\section{Alternative Expressions for Group Velocity}
It is essential to calculate the group velocity of each mode in
evaluating the transmission matrix. Although this velocity can be
computed using Eq.~(\ref{v}), we feel it more convenient to adopt an
alternative expression in numerical simulations. Below we derive this
expression. We shall first lay down the formulation for general lattices
and then specify it to ZGRs. 

Consider a general Bravais lattice, each site of which is labelled by
the corresponding position vector $\vec{R}$. Let the Hamiltonian be
\begin{equation}
\hat{H} = \sum_{\vec{R},\vec{R}'}H(\vec{R},\vec{R}')|\vec{R}\rangle\langle\vec{R}'|.
\end{equation}
We write the position operator of an electron as 
\begin{equation}
\hat{x} = \sum_{\vec{R}}\vec{R}|\vec{R}\rangle\langle\vec{R}|.
\end{equation}
The velocity operator is then defined to be 
\begin{equation}
\hat{V} = \dot{\hat{x}} = -
i[\vec{x},\hat{H}] = : \vec{v}(\vec{R},\vec{R}')|\vec{R}\rangle\langle\vec{R}'|,
\end{equation}
with
\begin{equation}
\vec{v}(\vec{R},\vec{R}') = -i\sum_{\vec{R},\vec{R}'}(\vec{R}-\vec{R}')H(\vec{R},\vec{R}').
\end{equation} 
Now the group velocity of any state $|\psi\rangle$ is given by
\begin{eqnarray}
\vec{v}_g = \langle\psi|\hat{V}|\psi\rangle =
 -i\sum_{\vec{R},\vec{R}'}\psi^*(\vec{R})(\vec{R}-\vec{R}')H(\vec{R},\vec{R}')\psi(\vec{R'}),
\label{vg}
\end{eqnarray}
where $\psi(\vec{R})$ denotes the wave function of the state.

Now we apply Eq.~(\ref{vg}) to ZGRs. We replace $\psi(\vec{R})$ by
Eq.~(\ref{propsi}) and after some algebra, we obtain the group velocity
(in the $x$-direction) assigned to the mode $\hat{\psi}({m,n;k,p,s})$ as follows,
\begin{equation}
v_{ps}(k) \propto -s(-1)^Q\sin\left(\frac{k}{2}\right)\sum_n\sin\left(p\left(N_z+1-n\right)\right)\sin\left(pn\right),
\end{equation}
up to an irrelevant constant. This expression does not involve a
derivative and is hence more convenient in numerical computations. The
group velocity vanishes in the $y$-direction expectedly.

\bibliographystyle{apsrev4-1}
\bibliography{dengbib}

\begin{thebibliography}{57}%
\makeatletter
\providecommand \@ifxundefined [1]{%
 \@ifx{#1\undefined}
}%
\providecommand \@ifnum [1]{%
 \ifnum #1\expandafter \@firstoftwo
 \else \expandafter \@secondoftwo
 \fi
}%
\providecommand \@ifx [1]{%
 \ifx #1\expandafter \@firstoftwo
 \else \expandafter \@secondoftwo
 \fi
}%
\providecommand \natexlab [1]{#1}%
\providecommand \enquote  [1]{``#1''}%
\providecommand \bibnamefont  [1]{#1}%
\providecommand \bibfnamefont [1]{#1}%
\providecommand \citenamefont [1]{#1}%
\providecommand \href@noop [0]{\@secondoftwo}%
\providecommand \href [0]{\begingroup \@sanitize@url \@href}%
\providecommand \@href[1]{\@@startlink{#1}\@@href}%
\providecommand \@@href[1]{\endgroup#1\@@endlink}%
\providecommand \@sanitize@url [0]{\catcode `\\12\catcode `\$12\catcode
  `\&12\catcode `\#12\catcode `\^12\catcode `\_12\catcode `\%12\relax}%
\providecommand \@@startlink[1]{}%
\providecommand \@@endlink[0]{}%
\providecommand \url  [0]{\begingroup\@sanitize@url \@url }%
\providecommand \@url [1]{\endgroup\@href {#1}{\urlprefix }}%
\providecommand \urlprefix  [0]{URL }%
\providecommand \Eprint [0]{\href }%
\providecommand \doibase [0]{http://dx.doi.org/}%
\providecommand \selectlanguage [0]{\@gobble}%
\providecommand \bibinfo  [0]{\@secondoftwo}%
\providecommand \bibfield  [0]{\@secondoftwo}%
\providecommand \translation [1]{[#1]}%
\providecommand \BibitemOpen [0]{}%
\providecommand \bibitemStop [0]{}%
\providecommand \bibitemNoStop [0]{.\EOS\space}%
\providecommand \EOS [0]{\spacefactor3000\relax}%
\providecommand \BibitemShut  [1]{\csname bibitem#1\endcsname}%
\let\auto@bib@innerbib\@empty
\bibitem [{\citenamefont {Novoselov}\ \emph {et~al.}(2004)\citenamefont
  {Novoselov}, \citenamefont {Geim}, \citenamefont {Morozov}, \citenamefont
  {Jiang}, \citenamefont {Zhang}, \citenamefont {Dubonos}, \citenamefont
  {Grigorieva},\ and\ \citenamefont {Firsov}}]{novoselov2004}%
  \BibitemOpen
  \bibfield  {author} {\bibinfo {author} {\bibfnamefont {K.~S.}\ \bibnamefont
  {Novoselov}}, \bibinfo {author} {\bibfnamefont {A.~K.}\ \bibnamefont {Geim}},
  \bibinfo {author} {\bibfnamefont {S.~V.}\ \bibnamefont {Morozov}}, \bibinfo
  {author} {\bibfnamefont {D.}~\bibnamefont {Jiang}}, \bibinfo {author}
  {\bibfnamefont {Y.}~\bibnamefont {Zhang}}, \bibinfo {author} {\bibfnamefont
  {S.~V.}\ \bibnamefont {Dubonos}}, \bibinfo {author} {\bibfnamefont {I.~V.}\
  \bibnamefont {Grigorieva}}, \ and\ \bibinfo {author} {\bibfnamefont {A.~A.}\
  \bibnamefont {Firsov}},\ }\href@noop {} {\bibfield  {journal} {\bibinfo
  {journal} {Science}\ }\textbf {\bibinfo {volume} {306}},\ \bibinfo {pages}
  {666} (\bibinfo {year} {2004})}\BibitemShut {NoStop}%
\bibitem [{\citenamefont {Novoselov}\ \emph
  {et~al.}(2005{\natexlab{a}})\citenamefont {Novoselov}, \citenamefont {Jiang},
  \citenamefont {Schedin}, \citenamefont {Booth}, \citenamefont {Khotkevich},
  \citenamefont {Morozov},\ and\ \citenamefont {Geim}}]{geim2005}%
  \BibitemOpen
  \bibfield  {author} {\bibinfo {author} {\bibfnamefont {K.}~\bibnamefont
  {Novoselov}}, \bibinfo {author} {\bibfnamefont {D.}~\bibnamefont {Jiang}},
  \bibinfo {author} {\bibfnamefont {F.}~\bibnamefont {Schedin}}, \bibinfo
  {author} {\bibfnamefont {T.}~\bibnamefont {Booth}}, \bibinfo {author}
  {\bibfnamefont {V.}~\bibnamefont {Khotkevich}}, \bibinfo {author}
  {\bibfnamefont {S.}~\bibnamefont {Morozov}}, \ and\ \bibinfo {author}
  {\bibfnamefont {A.}~\bibnamefont {Geim}},\ }\href@noop {} {\bibfield
  {journal} {\bibinfo  {journal} {Proc. Natl. Acad. Sci. USA}\ }\textbf
  {\bibinfo {volume} {102}},\ \bibinfo {pages} {10451} (\bibinfo {year}
  {2005}{\natexlab{a}})}\BibitemShut {NoStop}%
\bibitem [{\citenamefont {Novoselov}\ \emph
  {et~al.}(2005{\natexlab{b}})\citenamefont {Novoselov}, \citenamefont {Geim},
  \citenamefont {Morozov}, \citenamefont {Jiang}, \citenamefont {Katsnelson},
  \citenamefont {Grigorieva}, \citenamefont {Dubonos},\ and\ \citenamefont
  {Firsov}}]{novoselov2005b}%
  \BibitemOpen
  \bibfield  {author} {\bibinfo {author} {\bibfnamefont {K.}~\bibnamefont
  {Novoselov}}, \bibinfo {author} {\bibfnamefont {A.~K.}\ \bibnamefont {Geim}},
  \bibinfo {author} {\bibfnamefont {S.}~\bibnamefont {Morozov}}, \bibinfo
  {author} {\bibfnamefont {D.}~\bibnamefont {Jiang}}, \bibinfo {author}
  {\bibfnamefont {M.}~\bibnamefont {Katsnelson}}, \bibinfo {author}
  {\bibfnamefont {I.}~\bibnamefont {Grigorieva}}, \bibinfo {author}
  {\bibfnamefont {S.}~\bibnamefont {Dubonos}}, \ and\ \bibinfo {author}
  {\bibfnamefont {A.}~\bibnamefont {Firsov}},\ }\href@noop {} {\bibfield
  {journal} {\bibinfo  {journal} {Nature}\ }\textbf {\bibinfo {volume} {438}},\
  \bibinfo {pages} {197} (\bibinfo {year} {2005}{\natexlab{b}})}\BibitemShut
  {NoStop}%
\bibitem [{\citenamefont {Fujita}\ \emph {et~al.}(1996)\citenamefont {Fujita},
  \citenamefont {Wakabayashi}, \citenamefont {Nakada},\ and\ \citenamefont
  {Kusakabe}}]{fujita1996}%
  \BibitemOpen
  \bibfield  {author} {\bibinfo {author} {\bibfnamefont {M.}~\bibnamefont
  {Fujita}}, \bibinfo {author} {\bibfnamefont {K.}~\bibnamefont {Wakabayashi}},
  \bibinfo {author} {\bibfnamefont {K.}~\bibnamefont {Nakada}}, \ and\ \bibinfo
  {author} {\bibfnamefont {K.}~\bibnamefont {Kusakabe}},\ }\href@noop {}
  {\bibfield  {journal} {\bibinfo  {journal} {J. Phys. Soc. Jpn.}\ }\textbf
  {\bibinfo {volume} {65}},\ \bibinfo {pages} {1920} (\bibinfo {year}
  {1996})}\BibitemShut {NoStop}%
\bibitem [{\citenamefont {Nakada}\ \emph {et~al.}(1996)\citenamefont {Nakada},
  \citenamefont {Fujita}, \citenamefont {Dresselhaus},\ and\ \citenamefont
  {Dresselhaus}}]{nakada1996}%
  \BibitemOpen
  \bibfield  {author} {\bibinfo {author} {\bibfnamefont {K.}~\bibnamefont
  {Nakada}}, \bibinfo {author} {\bibfnamefont {M.}~\bibnamefont {Fujita}},
  \bibinfo {author} {\bibfnamefont {G.}~\bibnamefont {Dresselhaus}}, \ and\
  \bibinfo {author} {\bibfnamefont {M.~S.}\ \bibnamefont {Dresselhaus}},\
  }\href@noop {} {\bibfield  {journal} {\bibinfo  {journal} {Phys. Rev. B}\
  }\textbf {\bibinfo {volume} {54}},\ \bibinfo {pages} {17954} (\bibinfo {year}
  {1996})}\BibitemShut {NoStop}%
\bibitem [{\citenamefont {Wakabayashi}\ \emph {et~al.}(1999)\citenamefont
  {Wakabayashi}, \citenamefont {Fujita}, \citenamefont {Ajiki},\ and\
  \citenamefont {Sigrist}}]{wakabayashi1999}%
  \BibitemOpen
  \bibfield  {author} {\bibinfo {author} {\bibfnamefont {K.}~\bibnamefont
  {Wakabayashi}}, \bibinfo {author} {\bibfnamefont {M.}~\bibnamefont {Fujita}},
  \bibinfo {author} {\bibfnamefont {H.}~\bibnamefont {Ajiki}}, \ and\ \bibinfo
  {author} {\bibfnamefont {M.}~\bibnamefont {Sigrist}},\ }\href@noop {}
  {\bibfield  {journal} {\bibinfo  {journal} {Phys. Rev. B}\ }\textbf {\bibinfo
  {volume} {59}},\ \bibinfo {pages} {8271} (\bibinfo {year}
  {1999})}\BibitemShut {NoStop}%
\bibitem [{\citenamefont {Wakabayashi}\ \emph {et~al.}(1998)\citenamefont
  {Wakabayashi}, \citenamefont {Sigrist},\ and\ \citenamefont
  {Fujita}}]{wakabayashi1998jpsj}%
  \BibitemOpen
  \bibfield  {author} {\bibinfo {author} {\bibfnamefont {K.}~\bibnamefont
  {Wakabayashi}}, \bibinfo {author} {\bibfnamefont {M.}~\bibnamefont
  {Sigrist}}, \ and\ \bibinfo {author} {\bibfnamefont {M.}~\bibnamefont
  {Fujita}},\ }\href@noop {} {\bibfield  {journal} {\bibinfo  {journal} {J.
  Phys. Soc. Jpn.}\ }\textbf {\bibinfo {volume} {67}},\ \bibinfo {pages} {2089}
  (\bibinfo {year} {1998})}\BibitemShut {NoStop}%
\bibitem [{\citenamefont {Kane}\ and\ \citenamefont {Mele}(2005)}]{kane2005}%
  \BibitemOpen
  \bibfield  {author} {\bibinfo {author} {\bibfnamefont {C.~L.}\ \bibnamefont
  {Kane}}\ and\ \bibinfo {author} {\bibfnamefont {E.~J.}\ \bibnamefont
  {Mele}},\ }\href@noop {} {\bibfield  {journal} {\bibinfo  {journal} {Phys.
  Rev. Lett.}\ }\textbf {\bibinfo {volume} {95}},\ \bibinfo {pages} {146802}
  (\bibinfo {year} {2005})}\BibitemShut {NoStop}%
\bibitem [{\citenamefont {Wakabayashi}\ \emph {et~al.}(2007)\citenamefont
  {Wakabayashi}, \citenamefont {Takane},\ and\ \citenamefont
  {Sigrist}}]{wakabayashi2007}%
  \BibitemOpen
  \bibfield  {author} {\bibinfo {author} {\bibfnamefont {K.}~\bibnamefont
  {Wakabayashi}}, \bibinfo {author} {\bibfnamefont {Y.}~\bibnamefont {Takane}},
  \ and\ \bibinfo {author} {\bibfnamefont {M.}~\bibnamefont {Sigrist}},\
  }\href@noop {} {\bibfield  {journal} {\bibinfo  {journal} {Phys. Rev. Lett.}\
  }\textbf {\bibinfo {volume} {99}},\ \bibinfo {pages} {036601} (\bibinfo
  {year} {2007})}\BibitemShut {NoStop}%
\bibitem [{\citenamefont {Areshkin}\ \emph {et~al.}(2007)\citenamefont
  {Areshkin}, \citenamefont {Gunlycke},\ and\ \citenamefont
  {White}}]{white2007}%
  \BibitemOpen
  \bibfield  {author} {\bibinfo {author} {\bibfnamefont {D.~A.}\ \bibnamefont
  {Areshkin}}, \bibinfo {author} {\bibfnamefont {D.}~\bibnamefont {Gunlycke}},
  \ and\ \bibinfo {author} {\bibfnamefont {C.~T.}\ \bibnamefont {White}},\
  }\href@noop {} {\bibfield  {journal} {\bibinfo  {journal} {Nano Letters}\
  }\textbf {\bibinfo {volume} {7}},\ \bibinfo {pages} {204} (\bibinfo {year}
  {2007})}\BibitemShut {NoStop}%
\bibitem [{\citenamefont {Wakabayashi}\ \emph
  {et~al.}(2009{\natexlab{a}})\citenamefont {Wakabayashi}, \citenamefont
  {Takane}, \citenamefont {Yamamoto},\ and\ \citenamefont
  {Sigrist}}]{wakabayashi2009a}%
  \BibitemOpen
  \bibfield  {author} {\bibinfo {author} {\bibfnamefont {K.}~\bibnamefont
  {Wakabayashi}}, \bibinfo {author} {\bibfnamefont {Y.}~\bibnamefont {Takane}},
  \bibinfo {author} {\bibfnamefont {M.}~\bibnamefont {Yamamoto}}, \ and\
  \bibinfo {author} {\bibfnamefont {M.}~\bibnamefont {Sigrist}},\ }\href@noop
  {} {\bibfield  {journal} {\bibinfo  {journal} {Carbon}\ }\textbf {\bibinfo
  {volume} {47}},\ \bibinfo {pages} {124} (\bibinfo {year}
  {2009}{\natexlab{a}})}\BibitemShut {NoStop}%
\bibitem [{\citenamefont {Wakabayashi}\ \emph
  {et~al.}(2009{\natexlab{b}})\citenamefont {Wakabayashi}, \citenamefont
  {Takane}, \citenamefont {Yamamoto},\ and\ \citenamefont
  {Sigrist}}]{wakabayashi2009b}%
  \BibitemOpen
  \bibfield  {author} {\bibinfo {author} {\bibfnamefont {K.}~\bibnamefont
  {Wakabayashi}}, \bibinfo {author} {\bibfnamefont {Y.}~\bibnamefont {Takane}},
  \bibinfo {author} {\bibfnamefont {M.}~\bibnamefont {Yamamoto}}, \ and\
  \bibinfo {author} {\bibfnamefont {M.}~\bibnamefont {Sigrist}},\ }\href@noop
  {} {\bibfield  {journal} {\bibinfo  {journal} {New J. Phys.}\ }\textbf
  {\bibinfo {volume} {11}},\ \bibinfo {pages} {095016} (\bibinfo {year}
  {2009}{\natexlab{b}})}\BibitemShut {NoStop}%
\bibitem [{\citenamefont {Geim}\ and\ \citenamefont
  {Novoselov}(2007)}]{geim2007rise}%
  \BibitemOpen
  \bibfield  {author} {\bibinfo {author} {\bibfnamefont {A.~K.}\ \bibnamefont
  {Geim}}\ and\ \bibinfo {author} {\bibfnamefont {K.~S.}\ \bibnamefont
  {Novoselov}},\ }\href@noop {} {\bibfield  {journal} {\bibinfo  {journal}
  {Nat. Mater.}\ }\textbf {\bibinfo {volume} {6}},\ \bibinfo {pages} {183}
  (\bibinfo {year} {2007})}\BibitemShut {NoStop}%
\bibitem [{\citenamefont {Neto}\ \emph {et~al.}(2009)\citenamefont {Neto},
  \citenamefont {Guinea}, \citenamefont {Peres}, \citenamefont {Novoselov},\
  and\ \citenamefont {Geim}}]{neto2009electronic}%
  \BibitemOpen
  \bibfield  {author} {\bibinfo {author} {\bibfnamefont {A.~C.}\ \bibnamefont
  {Neto}}, \bibinfo {author} {\bibfnamefont {F.}~\bibnamefont {Guinea}},
  \bibinfo {author} {\bibfnamefont {N.}~\bibnamefont {Peres}}, \bibinfo
  {author} {\bibfnamefont {K.~S.}\ \bibnamefont {Novoselov}}, \ and\ \bibinfo
  {author} {\bibfnamefont {A.~K.}\ \bibnamefont {Geim}},\ }\href@noop {}
  {\bibfield  {journal} {\bibinfo  {journal} {Rev. Mod. Phys.}\ }\textbf
  {\bibinfo {volume} {81}},\ \bibinfo {pages} {109} (\bibinfo {year}
  {2009})}\BibitemShut {NoStop}%
\bibitem [{\citenamefont {Wakabayashi}\ and\ \citenamefont
  {Dutta}(2012)}]{wakabayashi2012nanoscale}%
  \BibitemOpen
  \bibfield  {author} {\bibinfo {author} {\bibfnamefont {K.}~\bibnamefont
  {Wakabayashi}}\ and\ \bibinfo {author} {\bibfnamefont {S.}~\bibnamefont
  {Dutta}},\ }\href@noop {} {\bibfield  {journal} {\bibinfo  {journal} {Solid
  State Comm.}\ }\textbf {\bibinfo {volume} {152}},\ \bibinfo {pages} {1420}
  (\bibinfo {year} {2012})}\BibitemShut {NoStop}%
\bibitem [{\citenamefont {Raza}(2012)}]{raza2012graphene}%
  \BibitemOpen
  \bibinfo {editor} {\bibfnamefont {H.}~\bibnamefont {Raza}},\ ed.,\ \href@noop
  {} {\emph {\bibinfo {title} {Graphene Nanoelectronics}}},\ Vol.~\bibinfo
  {volume} {1}\ (\bibinfo  {publisher} {Springer},\ \bibinfo {year}
  {2012})\BibitemShut {NoStop}%
\bibitem [{\citenamefont {Cheianov}\ and\ \citenamefont
  {Fal'ko}(2006)}]{cheianov2006}%
  \BibitemOpen
  \bibfield  {author} {\bibinfo {author} {\bibfnamefont {V.~V.}\ \bibnamefont
  {Cheianov}}\ and\ \bibinfo {author} {\bibfnamefont {V.~I.}\ \bibnamefont
  {Fal'ko}},\ }\href@noop {} {\bibfield  {journal} {\bibinfo  {journal} {Phys.
  Rev. B}\ }\textbf {\bibinfo {volume} {74}},\ \bibinfo {pages} {041403}
  (\bibinfo {year} {2006})}\BibitemShut {NoStop}%
\bibitem [{\citenamefont {Abanin}\ and\ \citenamefont
  {Levitov}(2007)}]{abanin2007}%
  \BibitemOpen
  \bibfield  {author} {\bibinfo {author} {\bibfnamefont {D.~A.}\ \bibnamefont
  {Abanin}}\ and\ \bibinfo {author} {\bibfnamefont {L.~S.}\ \bibnamefont
  {Levitov}},\ }\href@noop {} {\bibfield  {journal} {\bibinfo  {journal}
  {Science}\ }\textbf {\bibinfo {volume} {317}},\ \bibinfo {pages} {641}
  (\bibinfo {year} {2007})}\BibitemShut {NoStop}%
\bibitem [{\citenamefont {Deng}(2012)}]{deng2012}%
  \BibitemOpen
  \bibfield  {author} {\bibinfo {author} {\bibfnamefont {H.-Y.}\ \bibnamefont
  {Deng}},\ }\href@noop {} {\bibfield  {journal} {\bibinfo  {journal} {J. Appl.
  Phys.}\ }\textbf {\bibinfo {volume} {111}},\ \bibinfo {eid} {033706}
  (\bibinfo {year} {2012})}\BibitemShut {NoStop}%
\bibitem [{\citenamefont {Gunlycke}\ \emph {et~al.}(2007)\citenamefont
  {Gunlycke}, \citenamefont {Areshkin}, \citenamefont {Li}, \citenamefont
  {Mintmire},\ and\ \citenamefont {White}}]{gunlycke2007}%
  \BibitemOpen
  \bibfield  {author} {\bibinfo {author} {\bibfnamefont {D.}~\bibnamefont
  {Gunlycke}}, \bibinfo {author} {\bibfnamefont {D.~A.}\ \bibnamefont
  {Areshkin}}, \bibinfo {author} {\bibfnamefont {J.}~\bibnamefont {Li}},
  \bibinfo {author} {\bibfnamefont {J.~W.}\ \bibnamefont {Mintmire}}, \ and\
  \bibinfo {author} {\bibfnamefont {C.~T.}\ \bibnamefont {White}},\ }\href@noop
  {} {\bibfield  {journal} {\bibinfo  {journal} {Nano Letters}\ }\textbf
  {\bibinfo {volume} {7}},\ \bibinfo {pages} {3608} (\bibinfo {year}
  {2007})}\BibitemShut {NoStop}%
\bibitem [{\citenamefont {Areshkin}\ and\ \citenamefont
  {White}(2007)}]{areshkin2007}%
  \BibitemOpen
  \bibfield  {author} {\bibinfo {author} {\bibfnamefont {D.~A.}\ \bibnamefont
  {Areshkin}}\ and\ \bibinfo {author} {\bibfnamefont {C.~T.}\ \bibnamefont
  {White}},\ }\href@noop {} {\bibfield  {journal} {\bibinfo  {journal} {Nano
  Letters}\ }\textbf {\bibinfo {volume} {7}},\ \bibinfo {pages} {3253}
  (\bibinfo {year} {2007})}\BibitemShut {NoStop}%
\bibitem [{\citenamefont {\"Ozyilmaz}\ \emph {et~al.}(2007)\citenamefont
  {\"Ozyilmaz}, \citenamefont {Jarillo-Herrero}, \citenamefont {Efetov},
  \citenamefont {Abanin}, \citenamefont {Levitov},\ and\ \citenamefont
  {Kim}}]{ozyilmaz2007}%
  \BibitemOpen
  \bibfield  {author} {\bibinfo {author} {\bibfnamefont {B.}~\bibnamefont
  {\"Ozyilmaz}}, \bibinfo {author} {\bibfnamefont {P.}~\bibnamefont
  {Jarillo-Herrero}}, \bibinfo {author} {\bibfnamefont {D.}~\bibnamefont
  {Efetov}}, \bibinfo {author} {\bibfnamefont {D.~A.}\ \bibnamefont {Abanin}},
  \bibinfo {author} {\bibfnamefont {L.~S.}\ \bibnamefont {Levitov}}, \ and\
  \bibinfo {author} {\bibfnamefont {P.}~\bibnamefont {Kim}},\ }\href@noop {}
  {\bibfield  {journal} {\bibinfo  {journal} {Phys. Rev. Lett.}\ }\textbf
  {\bibinfo {volume} {99}},\ \bibinfo {pages} {166804} (\bibinfo {year}
  {2007})}\BibitemShut {NoStop}%
\bibitem [{\citenamefont {Huard}\ \emph {et~al.}(2007)\citenamefont {Huard},
  \citenamefont {Sulpizio}, \citenamefont {Stander}, \citenamefont {Todd},
  \citenamefont {Yang},\ and\ \citenamefont {Goldhaber-Gordon}}]{huard2007}%
  \BibitemOpen
  \bibfield  {author} {\bibinfo {author} {\bibfnamefont {B.}~\bibnamefont
  {Huard}}, \bibinfo {author} {\bibfnamefont {J.~A.}\ \bibnamefont {Sulpizio}},
  \bibinfo {author} {\bibfnamefont {N.}~\bibnamefont {Stander}}, \bibinfo
  {author} {\bibfnamefont {K.}~\bibnamefont {Todd}}, \bibinfo {author}
  {\bibfnamefont {B.}~\bibnamefont {Yang}}, \ and\ \bibinfo {author}
  {\bibfnamefont {D.}~\bibnamefont {Goldhaber-Gordon}},\ }\href@noop {}
  {\bibfield  {journal} {\bibinfo  {journal} {Phys. Rev. Lett.}\ }\textbf
  {\bibinfo {volume} {98}},\ \bibinfo {pages} {236803} (\bibinfo {year}
  {2007})}\BibitemShut {NoStop}%
\bibitem [{\citenamefont {Williams}\ \emph {et~al.}(2007)\citenamefont
  {Williams}, \citenamefont {DiCarlo},\ and\ \citenamefont
  {Marcus}}]{williams2007}%
  \BibitemOpen
  \bibfield  {author} {\bibinfo {author} {\bibfnamefont {J.~R.}\ \bibnamefont
  {Williams}}, \bibinfo {author} {\bibfnamefont {L.}~\bibnamefont {DiCarlo}}, \
  and\ \bibinfo {author} {\bibfnamefont {C.~M.}\ \bibnamefont {Marcus}},\
  }\href@noop {} {\bibfield  {journal} {\bibinfo  {journal} {Science}\ }\textbf
  {\bibinfo {volume} {317}},\ \bibinfo {pages} {638} (\bibinfo {year}
  {2007})}\BibitemShut {NoStop}%
\bibitem [{\citenamefont {Londergan}\ \emph {et~al.}(1999)\citenamefont
  {Londergan}, \citenamefont {Carini},\ and\ \citenamefont
  {Murdock}}]{londergan1999binding}%
  \BibitemOpen
  \bibfield  {author} {\bibinfo {author} {\bibfnamefont {J.~T.}\ \bibnamefont
  {Londergan}}, \bibinfo {author} {\bibfnamefont {J.~P.}\ \bibnamefont
  {Carini}}, \ and\ \bibinfo {author} {\bibfnamefont {D.~P.}\ \bibnamefont
  {Murdock}},\ }\href@noop {} {\emph {\bibinfo {title} {Binding and scattering
  in two-dimensional systems: applications to quantum wires, waveguides and
  photonic crystals}}},\ Vol.~\bibinfo {volume} {60}\ (\bibinfo  {publisher}
  {Springer},\ \bibinfo {year} {1999})\BibitemShut {NoStop}%
\bibitem [{\citenamefont {Landauer}(1957)}]{landauer1957spatial}%
  \BibitemOpen
  \bibfield  {author} {\bibinfo {author} {\bibfnamefont {R.}~\bibnamefont
  {Landauer}},\ }\href@noop {} {\bibfield  {journal} {\bibinfo  {journal} {IBM
  J. Res. Deve.}\ }\textbf {\bibinfo {volume} {1}},\ \bibinfo {pages} {223}
  (\bibinfo {year} {1957})}\BibitemShut {NoStop}%
\bibitem [{\citenamefont {Landauer}(1970)}]{landauer1970electrical}%
  \BibitemOpen
  \bibfield  {author} {\bibinfo {author} {\bibfnamefont {R.}~\bibnamefont
  {Landauer}},\ }\href@noop {} {\bibfield  {journal} {\bibinfo  {journal}
  {Philos. Mag.}\ }\textbf {\bibinfo {volume} {21}},\ \bibinfo {pages} {863}
  (\bibinfo {year} {1970})}\BibitemShut {NoStop}%
\bibitem [{\citenamefont {B\"uttiker}(1986)}]{buttiker1986}%
  \BibitemOpen
  \bibfield  {author} {\bibinfo {author} {\bibfnamefont {M.}~\bibnamefont
  {B\"uttiker}},\ }\href@noop {} {\bibfield  {journal} {\bibinfo  {journal}
  {Phys. Rev. Lett.}\ }\textbf {\bibinfo {volume} {57}},\ \bibinfo {pages}
  {1761} (\bibinfo {year} {1986})}\BibitemShut {NoStop}%
\bibitem [{\citenamefont {Ando}(1991)}]{ando1991quantum}%
  \BibitemOpen
  \bibfield  {author} {\bibinfo {author} {\bibfnamefont {T.}~\bibnamefont
  {Ando}},\ }\href@noop {} {\bibfield  {journal} {\bibinfo  {journal} {Phys.
  Rev. B}\ }\textbf {\bibinfo {volume} {44}},\ \bibinfo {pages} {8017}
  (\bibinfo {year} {1991})}\BibitemShut {NoStop}%
\bibitem [{\citenamefont {Zwierzycki}\ \emph {et~al.}(2008)\citenamefont
  {Zwierzycki}, \citenamefont {Khomyakov}, \citenamefont {Starikov},
  \citenamefont {Xia}, \citenamefont {Talanana}, \citenamefont {Xu},
  \citenamefont {Karpan}, \citenamefont {Marushchenko}, \citenamefont {Turek},
  \citenamefont {Bauer}, \citenamefont {Brocks},\ and\ \citenamefont
  {Kelly}}]{khomyakov2008}%
  \BibitemOpen
  \bibfield  {author} {\bibinfo {author} {\bibfnamefont {M.}~\bibnamefont
  {Zwierzycki}}, \bibinfo {author} {\bibfnamefont {P.~A.}\ \bibnamefont
  {Khomyakov}}, \bibinfo {author} {\bibfnamefont {A.~A.}\ \bibnamefont
  {Starikov}}, \bibinfo {author} {\bibfnamefont {K.}~\bibnamefont {Xia}},
  \bibinfo {author} {\bibfnamefont {M.}~\bibnamefont {Talanana}}, \bibinfo
  {author} {\bibfnamefont {P.~X.}\ \bibnamefont {Xu}}, \bibinfo {author}
  {\bibfnamefont {V.~M.}\ \bibnamefont {Karpan}}, \bibinfo {author}
  {\bibfnamefont {I.}~\bibnamefont {Marushchenko}}, \bibinfo {author}
  {\bibfnamefont {I.}~\bibnamefont {Turek}}, \bibinfo {author} {\bibfnamefont
  {G.~E.~W.}\ \bibnamefont {Bauer}}, \bibinfo {author} {\bibfnamefont
  {G.}~\bibnamefont {Brocks}}, \ and\ \bibinfo {author} {\bibfnamefont {P.~J.}\
  \bibnamefont {Kelly}},\ }\href {\doibase 10.1002/pssb.200743359} {\bibfield
  {journal} {\bibinfo  {journal} {physica status solidi (b)}\ }\textbf
  {\bibinfo {volume} {245}},\ \bibinfo {pages} {623} (\bibinfo {year}
  {2008})}\BibitemShut {NoStop}%
\bibitem [{\citenamefont {S{\o}rensen}\ \emph {et~al.}(2009)\citenamefont
  {S{\o}rensen}, \citenamefont {Hansen}, \citenamefont {Petersen},
  \citenamefont {Skelboe},\ and\ \citenamefont
  {Stokbro}}]{sorensen2009efficient}%
  \BibitemOpen
  \bibfield  {author} {\bibinfo {author} {\bibfnamefont {H.~H.~B.}\
  \bibnamefont {S{\o}rensen}}, \bibinfo {author} {\bibfnamefont {P.~C.}\
  \bibnamefont {Hansen}}, \bibinfo {author} {\bibfnamefont {D.~E.}\
  \bibnamefont {Petersen}}, \bibinfo {author} {\bibfnamefont {S.}~\bibnamefont
  {Skelboe}}, \ and\ \bibinfo {author} {\bibfnamefont {K.}~\bibnamefont
  {Stokbro}},\ }\href@noop {} {\bibfield  {journal} {\bibinfo  {journal} {Phys.
  Rev. B}\ }\textbf {\bibinfo {volume} {79}},\ \bibinfo {pages} {205322}
  (\bibinfo {year} {2009})}\BibitemShut {NoStop}%
\bibitem [{\citenamefont {Economou}\ and\ \citenamefont
  {Soukoulis}(1981)}]{economou1981}%
  \BibitemOpen
  \bibfield  {author} {\bibinfo {author} {\bibfnamefont {E.~N.}\ \bibnamefont
  {Economou}}\ and\ \bibinfo {author} {\bibfnamefont {C.~M.}\ \bibnamefont
  {Soukoulis}},\ }\href {\doibase 10.1103/PhysRevLett.46.618} {\bibfield
  {journal} {\bibinfo  {journal} {Phys. Rev. Lett.}\ }\textbf {\bibinfo
  {volume} {46}},\ \bibinfo {pages} {618} (\bibinfo {year} {1981})}\BibitemShut
  {NoStop}%
\bibitem [{\citenamefont {Fisher}\ and\ \citenamefont
  {Lee}(1981)}]{fisher1981relation}%
  \BibitemOpen
  \bibfield  {author} {\bibinfo {author} {\bibfnamefont {D.~S.}\ \bibnamefont
  {Fisher}}\ and\ \bibinfo {author} {\bibfnamefont {P.~A.}\ \bibnamefont
  {Lee}},\ }\href@noop {} {\bibfield  {journal} {\bibinfo  {journal} {Phys.
  Rev. B}\ }\textbf {\bibinfo {volume} {23}},\ \bibinfo {pages} {6851}
  (\bibinfo {year} {1981})}\BibitemShut {NoStop}%
\bibitem [{\citenamefont {Datta}(1997)}]{datta1997electronic}%
  \BibitemOpen
  \bibfield  {author} {\bibinfo {author} {\bibfnamefont {S.}~\bibnamefont
  {Datta}},\ }\href@noop {} {\emph {\bibinfo {title} {Electronic transport in
  mesoscopic systems}}}\ (\bibinfo  {publisher} {Cambridge university press},\
  \bibinfo {year} {1997})\BibitemShut {NoStop}%
\bibitem [{\citenamefont {Khomyakov}\ \emph {et~al.}(2005)\citenamefont
  {Khomyakov}, \citenamefont {Brocks}, \citenamefont {Karpan}, \citenamefont
  {Zwierzycki},\ and\ \citenamefont {Kelly}}]{khomyakov2005conductance}%
  \BibitemOpen
  \bibfield  {author} {\bibinfo {author} {\bibfnamefont {P.}~\bibnamefont
  {Khomyakov}}, \bibinfo {author} {\bibfnamefont {G.}~\bibnamefont {Brocks}},
  \bibinfo {author} {\bibfnamefont {V.}~\bibnamefont {Karpan}}, \bibinfo
  {author} {\bibfnamefont {M.}~\bibnamefont {Zwierzycki}}, \ and\ \bibinfo
  {author} {\bibfnamefont {P.}~\bibnamefont {Kelly}},\ }\href@noop {}
  {\bibfield  {journal} {\bibinfo  {journal} {Phys. Rev. B}\ }\textbf {\bibinfo
  {volume} {72}},\ \bibinfo {pages} {035450} (\bibinfo {year}
  {2005})}\BibitemShut {NoStop}%
\bibitem [{\citenamefont {Wurm}\ \emph {et~al.}(2009)\citenamefont {Wurm},
  \citenamefont {Wimmer}, \citenamefont {Adagideli}, \citenamefont {Richter},\
  and\ \citenamefont {Baranger}}]{wurm2009interfaces}%
  \BibitemOpen
  \bibfield  {author} {\bibinfo {author} {\bibfnamefont {J.}~\bibnamefont
  {Wurm}}, \bibinfo {author} {\bibfnamefont {M.}~\bibnamefont {Wimmer}},
  \bibinfo {author} {\bibfnamefont {{\.I}.}~\bibnamefont {Adagideli}}, \bibinfo
  {author} {\bibfnamefont {K.}~\bibnamefont {Richter}}, \ and\ \bibinfo
  {author} {\bibfnamefont {H.~U.}\ \bibnamefont {Baranger}},\ }\href@noop {}
  {\bibfield  {journal} {\bibinfo  {journal} {New J. Phys.}\ }\textbf {\bibinfo
  {volume} {11}},\ \bibinfo {pages} {095022} (\bibinfo {year}
  {2009})}\BibitemShut {NoStop}%
\bibitem [{\citenamefont {Iyengar}\ \emph {et~al.}(2008)\citenamefont
  {Iyengar}, \citenamefont {Luo}, \citenamefont {Fertig},\ and\ \citenamefont
  {Brey}}]{iyengar2008}%
  \BibitemOpen
  \bibfield  {author} {\bibinfo {author} {\bibfnamefont {A.}~\bibnamefont
  {Iyengar}}, \bibinfo {author} {\bibfnamefont {T.}~\bibnamefont {Luo}},
  \bibinfo {author} {\bibfnamefont {H.~A.}\ \bibnamefont {Fertig}}, \ and\
  \bibinfo {author} {\bibfnamefont {L.}~\bibnamefont {Brey}},\ }\href {\doibase
  10.1103/PhysRevB.78.235411} {\bibfield  {journal} {\bibinfo  {journal} {Phys.
  Rev. B}\ }\textbf {\bibinfo {volume} {78}},\ \bibinfo {pages} {235411}
  (\bibinfo {year} {2008})}\BibitemShut {NoStop}%
\bibitem [{\citenamefont {Brey}\ and\ \citenamefont {Fertig}(2006)}]{brey2006}%
  \BibitemOpen
  \bibfield  {author} {\bibinfo {author} {\bibfnamefont {L.}~\bibnamefont
  {Brey}}\ and\ \bibinfo {author} {\bibfnamefont {H.~A.}\ \bibnamefont
  {Fertig}},\ }\href {\doibase 10.1103/PhysRevB.73.235411} {\bibfield
  {journal} {\bibinfo  {journal} {Phys. Rev. B}\ }\textbf {\bibinfo {volume}
  {73}},\ \bibinfo {pages} {235411} (\bibinfo {year} {2006})}\BibitemShut
  {NoStop}%
\bibitem [{\citenamefont {Beenakker}(1997)}]{beenakker1997}%
  \BibitemOpen
  \bibfield  {author} {\bibinfo {author} {\bibfnamefont {C.~W.~J.}\
  \bibnamefont {Beenakker}},\ }\href {\doibase 10.1103/RevModPhys.69.731}
  {\bibfield  {journal} {\bibinfo  {journal} {Rev. Mod. Phys.}\ }\textbf
  {\bibinfo {volume} {69}},\ \bibinfo {pages} {731} (\bibinfo {year}
  {1997})}\BibitemShut {NoStop}%
\bibitem [{\citenamefont {Yamamoto}\ and\ \citenamefont
  {Wakabayashi}(2009)}]{yamamoto2009}%
  \BibitemOpen
  \bibfield  {author} {\bibinfo {author} {\bibfnamefont {M.}~\bibnamefont
  {Yamamoto}}\ and\ \bibinfo {author} {\bibfnamefont {K.}~\bibnamefont
  {Wakabayashi}},\ }\href@noop {} {\bibfield  {journal} {\bibinfo  {journal}
  {Appl. Phys. Lett.}\ }\textbf {\bibinfo {volume} {95}},\ \bibinfo {eid}
  {082109} (\bibinfo {year} {2009})}\BibitemShut {NoStop}%
\bibitem [{\citenamefont {Rycerz}\ \emph {et~al.}(2007)\citenamefont {Rycerz},
  \citenamefont {Tworzyd{\l}o},\ and\ \citenamefont {Beenakker}}]{rycerz2007}%
  \BibitemOpen
  \bibfield  {author} {\bibinfo {author} {\bibfnamefont {A.}~\bibnamefont
  {Rycerz}}, \bibinfo {author} {\bibfnamefont {J.}~\bibnamefont
  {Tworzyd{\l}o}}, \ and\ \bibinfo {author} {\bibfnamefont {C.}~\bibnamefont
  {Beenakker}},\ }\href@noop {} {\bibfield  {journal} {\bibinfo  {journal}
  {Nat. Phys.}\ }\textbf {\bibinfo {volume} {3}},\ \bibinfo {pages} {172}
  (\bibinfo {year} {2007})}\BibitemShut {NoStop}%
\bibitem [{\citenamefont {Wakabayashi}\ and\ \citenamefont
  {Sigrist}(2000)}]{wakabayashi2000}%
  \BibitemOpen
  \bibfield  {author} {\bibinfo {author} {\bibfnamefont {K.}~\bibnamefont
  {Wakabayashi}}\ and\ \bibinfo {author} {\bibfnamefont {M.}~\bibnamefont
  {Sigrist}},\ }\href {\doibase 10.1103/PhysRevLett.84.3390} {\bibfield
  {journal} {\bibinfo  {journal} {Phys. Rev. Lett.}\ }\textbf {\bibinfo
  {volume} {84}},\ \bibinfo {pages} {3390} (\bibinfo {year}
  {2000})}\BibitemShut {NoStop}%
\bibitem [{In numerical scheme()}]{note3}%
  \BibitemOpen
  In numerical scheme,\ \href@noop {} {}\bibinfo {note} {Eq.~(\ref{Q}) can lead
  to ambiguity in calculating the parity. Take propagating modes for example,
  i.e., $\lambda = e^{ik}$. The parity does not depend on $k$ and can be
  evaluated at e.g. $k=\pi$, where $\Lambda = 0$ and $p=\frac{\pi}{N_z}\cdot
  r$, with $r = 1,...,N_z$ being an integer. Hence, $(-1)^Q = (-1)^r\cdot
  \sigma^{-1}\sqrt{\sigma^2} = (-1)^r\cdot \sigma^{-1}\cdot (\pm\sigma) = \pm
  \cdot (-1)^r$, where the amiguity about the sign can be removed by fixing it
  to be plus. The computer, however, may be ill in doing this and can produce
  messy results. Therefore, it is crucial to ensure that $\sqrt{\sigma^2} =
  \sigma$ in programming.}\BibitemShut {Stop}%
\bibitem [{The $\varepsilon_{max}$ occurs at $k=0$()}]{note2}%
  \BibitemOpen
  The $\varepsilon_{max}$ occurs at $k=0$,\ \href@noop {} {}\bibinfo {note} {~
  i.e., $\Lambda=2$. Then, by Eq.~(\ref{E}), $\varepsilon^2_{max} =
  5+4\cos(\phi)$. The $\phi$ plays the role in
  $\cos(\phi)=-\frac{1}{2}-\frac{\sin(\phi)}{\tan(N_z\phi)}$, as inferred from
  Eq.~(\ref{lam}). Assume $\phi=\frac{\pi}{N_z}-\delta$, where
  $0<\delta\ll\frac{\pi}{N_z}$, and we have
  $\cos(\frac{\pi}{N_z}-\delta)=\frac{\sin(\frac{\pi}{N_z}-\delta)}{\tan(N_z\delta)}-\frac{1}{2}$.
  For very large $N_z$, this translates into $\delta \approx
  \frac{2\pi}{N_z(3N_z+2)}$, which is indeed much smaller than
  $\frac{\pi}{N_z}$ as presumed. Upon subtitution, we arrive at the
  $\varepsilon_{max}$ given in the main text.}\BibitemShut {Stop}%
\bibitem [{To prove this()}]{note1}%
  \BibitemOpen
  To prove this,\ \href@noop {} {}\bibinfo {note} {~one may show that the
  companion matrix of the polynomial is irreducible. Note that this polynomial
  can be written as $P(x)=\sum^{2N_z}_{j=0}C_jx^j$, where $C_j=1+j$ for
  $j<N_z$, $C_j=N_z+1-\frac{1}{\varepsilon^2}$ for $j=N_z$ and $C_j=2N_z+1-j$
  for $j>N_z$.}\BibitemShut {Stop}%
\bibitem [{\citenamefont {Akhmerov}\ \emph {et~al.}(2008)\citenamefont
  {Akhmerov}, \citenamefont {Bardarson}, \citenamefont {Rycerz},\ and\
  \citenamefont {Beenakker}}]{akhmerov2008}%
  \BibitemOpen
  \bibfield  {author} {\bibinfo {author} {\bibfnamefont {A.}~\bibnamefont
  {Akhmerov}}, \bibinfo {author} {\bibfnamefont {J.}~\bibnamefont {Bardarson}},
  \bibinfo {author} {\bibfnamefont {A.}~\bibnamefont {Rycerz}}, \ and\ \bibinfo
  {author} {\bibfnamefont {C.}~\bibnamefont {Beenakker}},\ }\href@noop {}
  {\bibfield  {journal} {\bibinfo  {journal} {Phys. Rev. B}\ }\textbf {\bibinfo
  {volume} {77}},\ \bibinfo {pages} {205416} (\bibinfo {year}
  {2008})}\BibitemShut {NoStop}%
\bibitem [{\citenamefont {Wakabayashi}\ and\ \citenamefont
  {Aoki}(2002)}]{wakabayashi2002intb}%
  \BibitemOpen
  \bibfield  {author} {\bibinfo {author} {\bibfnamefont {K.}~\bibnamefont
  {Wakabayashi}}\ and\ \bibinfo {author} {\bibfnamefont {T.}~\bibnamefont
  {Aoki}},\ }\href@noop {} {\bibfield  {journal} {\bibinfo  {journal} {Int. J.
  Mod. Phys. B}\ }\textbf {\bibinfo {volume} {16}},\ \bibinfo {pages} {4897}
  (\bibinfo {year} {2002})}\BibitemShut {NoStop}%
\bibitem [{\citenamefont {Deng}\ \emph {et~al.}(2013)\citenamefont {Deng},
  \citenamefont {Wakabayashi},\ and\ \citenamefont {Lam}}]{deng2013jpsj}%
  \BibitemOpen
  \bibfield  {author} {\bibinfo {author} {\bibfnamefont {H.-Y.}\ \bibnamefont
  {Deng}}, \bibinfo {author} {\bibfnamefont {K.}~\bibnamefont {Wakabayashi}}, \
  and\ \bibinfo {author} {\bibfnamefont {C.-H.}\ \bibnamefont {Lam}},\ }\href
  {\doibase 10.7566/JPSJ.82.104707} {\bibfield  {journal} {\bibinfo  {journal}
  {J. Phys. Soc. Jpn.}\ }\textbf {\bibinfo {volume} {82}},\ \bibinfo {pages}
  {104707} (\bibinfo {year} {2013})}\BibitemShut {NoStop}%
\bibitem [{\citenamefont {Kerner}(1966)}]{kerner1966gesamtschrittverfahren}%
  \BibitemOpen
  \bibfield  {author} {\bibinfo {author} {\bibfnamefont {I.~O.}\ \bibnamefont
  {Kerner}},\ }\href@noop {} {\bibfield  {journal} {\bibinfo  {journal} {Numer.
  Math.}\ }\textbf {\bibinfo {volume} {8}},\ \bibinfo {pages} {290} (\bibinfo
  {year} {1966})}\BibitemShut {NoStop}%
\bibitem [{\citenamefont {Durand}(1972)}]{durand1972solutions}%
  \BibitemOpen
  \bibfield  {author} {\bibinfo {author} {\bibfnamefont {{\'E}.}~\bibnamefont
  {Durand}},\ }\href@noop {} {\bibfield  {journal} {\bibinfo  {journal} {Paris:
  Masson, 1972}\ }\textbf {\bibinfo {volume} {1}} (\bibinfo {year}
  {1972})}\BibitemShut {NoStop}%
\bibitem [{\citenamefont {Klein}(1994)}]{klein1994graphitic}%
  \BibitemOpen
  \bibfield  {author} {\bibinfo {author} {\bibfnamefont {D.}~\bibnamefont
  {Klein}},\ }\href@noop {} {\bibfield  {journal} {\bibinfo  {journal} {Chem.
  Phys. Lett.}\ }\textbf {\bibinfo {volume} {217}},\ \bibinfo {pages} {261}
  (\bibinfo {year} {1994})}\BibitemShut {NoStop}%
\bibitem [{\citenamefont {Wakabayashi}(2001)}]{wakabayashi2001prb}%
  \BibitemOpen
  \bibfield  {author} {\bibinfo {author} {\bibfnamefont {K.}~\bibnamefont
  {Wakabayashi}},\ }\href@noop {} {\bibfield  {journal} {\bibinfo  {journal}
  {Phys. Rev. B}\ }\textbf {\bibinfo {volume} {64}},\ \bibinfo {pages} {125428}
  (\bibinfo {year} {2001})}\BibitemShut {NoStop}%
\bibitem [{\citenamefont {Deng}\ \emph {et~al.}(2014)\citenamefont {Deng},
  \citenamefont {Wakabayashi},\ and\ \citenamefont {Lam}}]{deng2014prb}%
  \BibitemOpen
  \bibfield  {author} {\bibinfo {author} {\bibfnamefont {H.-Y.}\ \bibnamefont
  {Deng}}, \bibinfo {author} {\bibfnamefont {K.}~\bibnamefont {Wakabayashi}}, \
  and\ \bibinfo {author} {\bibfnamefont {C.-H.}\ \bibnamefont {Lam}},\ }\href
  {\doibase 10.1103/PhysRevB.89.045423} {\bibfield  {journal} {\bibinfo
  {journal} {Phys. Rev. B}\ }\textbf {\bibinfo {volume} {89}},\ \bibinfo
  {pages} {045423} (\bibinfo {year} {2014})}\BibitemShut {NoStop}%
\bibitem [{\citenamefont {Caroli}\ \emph {et~al.}(1971)\citenamefont {Caroli},
  \citenamefont {Combescot}, \citenamefont {Nozieres},\ and\ \citenamefont
  {Saint-James}}]{caroli1971direct}%
  \BibitemOpen
  \bibfield  {author} {\bibinfo {author} {\bibfnamefont {C.}~\bibnamefont
  {Caroli}}, \bibinfo {author} {\bibfnamefont {R.}~\bibnamefont {Combescot}},
  \bibinfo {author} {\bibfnamefont {P.}~\bibnamefont {Nozieres}}, \ and\
  \bibinfo {author} {\bibfnamefont {D.}~\bibnamefont {Saint-James}},\
  }\href@noop {} {\bibfield  {journal} {\bibinfo  {journal} {J. Phys. C: Solid
  State}\ }\textbf {\bibinfo {volume} {4}},\ \bibinfo {pages} {916} (\bibinfo
  {year} {1971})}\BibitemShut {NoStop}%
\bibitem [{\citenamefont {Ferreira}\ \emph {et~al.}(2011)\citenamefont
  {Ferreira}, \citenamefont {Leuenberger}, \citenamefont {Loss},\ and\
  \citenamefont {Egues}}]{ferreira2011}%
  \BibitemOpen
  \bibfield  {author} {\bibinfo {author} {\bibfnamefont {G.~J.}\ \bibnamefont
  {Ferreira}}, \bibinfo {author} {\bibfnamefont {M.~N.}\ \bibnamefont
  {Leuenberger}}, \bibinfo {author} {\bibfnamefont {D.}~\bibnamefont {Loss}}, \
  and\ \bibinfo {author} {\bibfnamefont {J.~C.}\ \bibnamefont {Egues}},\
  }\href@noop {} {\bibfield  {journal} {\bibinfo  {journal} {Phys. Rev. B}\
  }\textbf {\bibinfo {volume} {84}},\ \bibinfo {pages} {125453} (\bibinfo
  {year} {2011})}\BibitemShut {NoStop}%
\bibitem [{\citenamefont {Hern\'{a}ndez}\ and\ \citenamefont
  {Lewenkopf}(2012)}]{hernandez2012}%
  \BibitemOpen
  \bibfield  {author} {\bibinfo {author} {\bibfnamefont {A.~R.}\ \bibnamefont
  {Hern\'{a}ndez}}\ and\ \bibinfo {author} {\bibfnamefont {C.~H.}\ \bibnamefont
  {Lewenkopf}},\ }\href@noop {} {\bibfield  {journal} {\bibinfo  {journal}
  {Phys. Rev. B}\ }\textbf {\bibinfo {volume} {86}},\ \bibinfo {pages} {155439}
  (\bibinfo {year} {2012})}\BibitemShut {NoStop}%
\bibitem [{\citenamefont {Grover}\ \emph {et~al.}(2010)\citenamefont {Grover},
  \citenamefont {Ghosh},\ and\ \citenamefont {Sharma}}]{grover2012}%
  \BibitemOpen
  \bibfield  {author} {\bibinfo {author} {\bibfnamefont {S.}~\bibnamefont
  {Grover}}, \bibinfo {author} {\bibfnamefont {S.}~\bibnamefont {Ghosh}}, \
  and\ \bibinfo {author} {\bibfnamefont {M.}~\bibnamefont {Sharma}},\
  }\href@noop {} {\bibfield  {journal} {\bibinfo  {journal} {Modelling Simul.
  Mater. Sci. Eng.}\ }\textbf {\bibinfo {volume} {20}},\ \bibinfo {pages}
  {045010} (\bibinfo {year} {2010})}\BibitemShut {NoStop}%
\end{thebibliography}%

\end{document}